\documentclass[a4paper,fleqn,usenatbib]{mnras}
\usepackage{color}
\usepackage[T1]{fontenc}
\usepackage{amsmath}	% Advanced maths commands
\usepackage{amssymb}	% Extra maths symbols
\usepackage{ae,aecompl}
\usepackage{graphicx}
\usepackage{rotating}
\usepackage{lscape}
\usepackage{amstext}
\usepackage{multirow}
\usepackage{graphicx}
\usepackage{hyperref}
\usepackage{caption}

\title[Geometry of the Large Magellanic Cloud]
{Geometry of the Large Magellanic Cloud Using Multi- wavelength Photometry of 
Classical Cepheids}
\author[Sukanta Deb et al.]{Sukanta Deb$^{1,2}$\thanks{E-mail: sukanta.deb@cottonuniversity.ac.in},
Chow-Choong Ngeow$^{3}$, Shashi M. Kanbur$^{4}$, Harinder P. Singh$^{5}$, 
\newauthor Daniel Wysocki$^{6}$, Subhash Kumar$^{7}$ \\
$^{1}$Department of Physics, Cotton University, Panbazar, Guwahati 781001,
Assam, India \\
$^{2}$Space and Astronomy Research Center, Cotton University, Panbazar, Guwahati 781001, Assam, India \\
$^{3}$Graduate Institute of Astronomy, National Central University, Jhongli 
32001, Taiwan \\ 
$^4$Department of Physics, State University of New York at Oswego, Oswego, 
NY 13126, USA \\
$^{5}$Department of Physics \& Astrophysics, University of Delhi, Delhi 
110007, India \\
$^{6}$Center for Computational Relativity and Gravitation, Rochester Institute 
of Technology, Rochester, NY 14623, USA \\ 
$^{7}$Department of Physics, Acharya Narendra Dev College, Govindpuri, Kalkaji, New Delhi 110019, India 
}
\date{Received on ; Accepted on }
\pubyear{2017}
\begin{document}
\label{firstpage}
\pagerange{\pageref{firstpage}--\pageref{lastpage}}
\maketitle
\begin{abstract}
We determine the geometrical and viewing angle parameters of the Large 
Magellanic Cloud (LMC) using the Leavitt law based on a sample of more than 
$3500$ common classical Cepheids (FU and FO) in optical ($V,I$), 
near-infrared ($JHK_{s}$) and mid-infrared ($[3.6]~\mu$m and $[4.5]~\mu$m) 
photometric bands. Statistical reddening and distance modulus free from the 
effect of reddening to each of the individual Cepheids are obtained  using the 
simultaneous multi-band fit to the apparent distance moduli from the analysis 
of the resulting Leavitt laws in these seven photometric bands. A reddening 
map of the LMC obtained from the analysis shows good agreement with the other 
maps available in the literature. Extinction free distance measurements along 
with the information of the equatorial coordinates $(\alpha,\delta)$ for 
individual stars are used to obtain the corresponding Cartesian coordinates 
with respect to the plane of the sky. By fitting a plane solution of the form 
$z=f(x,y)$ to the observed three dimensional distribution, the following 
viewing angle parameters of the LMC are obtained: inclination angle 
$i=25^{\circ}.110\pm 0^{\circ}.365$, position angle of line of nodes 
$\theta_{\text{lon}}=154^{\circ}.702\pm1^{\circ}.378$.  
On the other hand, modelling the observed three dimensional distribution of 
the Cepheids as a triaxial ellipsoid, the following values of the geometrical 
axes ratios of the LMC are obtained: $1.000\pm 0.003:1.151\pm0.003:1.890\pm 0.014$   with the viewing angle parameters:  inclination angle of 
$i=11^{\circ}.920\pm 0^{\circ}.315$ with respect to the longest axis from the 
line of sight and position angle of line of nodes $\theta_{\rm lon} = 128^{\circ}.871\pm 0^{\circ}.569$. The position angles are measured eastwards from 
north.    
\end{abstract}

\section{Introduction}
The LMC is classified as a barred spiral galaxy of type SB(s)m, the prototype 
of a class of Magellanic spirals \citep{deva72,deva91}. The galaxy is 
characterised by a prominent offset stellar bar located near its  center 
with the dominant spiral arm to the north with two ``embryonic" arms situated 
to the south. 
The 
galaxy is situated at a distance of $49.97\pm0.19(\text{stat.})\pm 1.11(\text{sys.})$ kpc \citep{piet13}. Because of its proximity to the Milky Way, the 
LMC offers an excellent laboratory to address many important 
astrophysical issues such as star formation, morphological structure, dust 
distribution using various tracers, microlensing observed towards the LMC, 
gravitational and hydrodynamical interaction as well as structure formation 
and evolution of galaxies. The accurate distance determination to the LMC 
plays a pivotal role in calibrating the extragalactic distance scale. Over the 
past two decades, considerable progress has been made in reducing the 
systematic uncertainty in the LMC distance determination in order to increase 
the accuracy of the value of the Hubble constant $H_{0}.$

Cepheids have played a vital role in the extragalactic distance determinations 
in astrophysics since the discovery of the period-luminosity (PL) relations 
called the Leavitt law 
\citep{leav08,leav12}. 
Since Cepheids are 
young stars, they are found in abundance in a spiral galaxy such as the LMC 
\citep{free01}. Because of negligible values of interstellar extinction 
(reduced by a factor of $\sim 20$ compared to optical wavelengths)  
and reduced sensitivity to metallicity, longer wavelength Cepheid data have 
been utilized in a large number of studies to calibrate the 
LMC Cepheid PL relations for distance determinations: 
outlined in \citet{chow09_aip}, \citet{free11}, \citet{bhar16a}, 
\citet{bhar16} and \citet{bhar17}. Using the 
mid-infrared observations of $Spitzer$, a large number of classical Cepheid 
data were collected by SAGE 
(Surveying the Agents of a Galaxy's Evolution) project \citep{meix06}. 
Based on these single epoch observations, 
\citet{chow08,free08}, among others,  derived mid-infrared PL relations 
for LMC fundamental mode Cepheids at four 
wavelength bands.

The use of multi-wavelength PL relations of Cepheids in the extra-galactic 
distance-scale studies has a number of advantages: (i) Distance 
determinations are less affected due to the differences in chemical composition
from one galaxy to the other/among different regions of the same galaxy/among 
different photometric bands and hence consistent values of distances are 
possible, (ii) Dependence of zero points of the Cepheid PL relations on 
metallicity can be minimised once corrected for reddening using 
multi-wavelength PL relations and hence the systematic error in distance 
determination can be reduced significantly, (iii) Independent determinations of 
true distance modulus and reddening \citep{free90,free01}. 
However, the extinction corrections due to interstellar reddening rely on the 
assumption of a universal behaviour of Cepheids at different wavelengths and a 
universality of the Galactic extinction law \citep{free01}. The independent 
reddening values obtained from the multi-wavelength PL relations are  
crucial to construct reddening map of the host galaxy which is 
important to study the dust distribution among different regions of the host 
galaxy. 
Furthermore, the use of 
multi-wavelength photometry using more number of bands not only increases the 
figure of merit ratio but also takes into account complex interplay among 
magnitudes in different photometric bands. This helps in disentangling part of 
magnitude variations due to distance and other factors like age, metallicity, 
reddening, etc., \citep{niko04}.                         
       
The road map of the present investigation using the multi-wavelength archival 
data of common LMC classical Cepheids in seven photometric bands, 
viz., $V,I,J,H,K$,~$[3.6]~ \mu m $ and $[4.5]~\mu m$  consists of the 
following : 
i) new improved PL relations of classical Cepheids using the \citet{niko04} 
method, ii) independent determinations of accurate values of reddening and 
reddening-corrected distances to the individual classical Cepheids with 
respect to the mean reddening and distance of the LMC from the 
simultaneous solutions of multi-band data, iii) highly precise and accurate 
reddening map as well as geometrical and viewing angle parameters of the LMC,
iv) Separation of bar and disk of the LMC as well as determinations of the 
viewing angle parameters for these two distinct structures using the projected 
Cartesian coordinates. The Cartesian coordinates and the distance values 
obtained from the multi-wavelength analysis also facilitate the comparison of 
the line of sight distances among different regions of the LMC disk based on a 
statistical test.    
 
One of the important reasons for using the data only for the mid-infrared 
$[3.6]~\mu m$ and $[4.5]~\mu m$ photometric bands from the existing four 
mid-infrared bands is that the magnitude values are available for a statistically large 
sample of classical Cepheids in these two photometric  bands from the SAGE 
catalog corresponding to the common stars with  $V,I, J,H, Ks$ observations 
from \citet{inno16}. The catalog compiled by \citet{inno16} contains the 
largest data set for classical Cepheids in the near infrared based on the 
OGLE-IV catalog \citep{sosz15}. The near-infrared mean magnitudes were 
compiled in \citet{inno16} using the data from the VMC survey, LMC NIR 
Synoptic survey, IRSF sample, 2MASS catalog as well as from \citet{pers04}. Mostly, the near-infrared mean magnitudes not available in those databases were 
determined  in the catalog based on the template fitting technique developed by 
\citet{inno15}, whereas the determinations of the OGLE-IV optical mean 
magnitudes were based on the seventh order Fourier-series fit. Further details 
of the LMC Cepheid catalog compilation can be found in \citet{inno16}.

The present sample of  classical Cepheids is restricted to those stars which 
have the complementary data in all seven wavelength photometric bands. 
One of the first studies dealing with the determination of the geometry of the 
disk of the LMC using a large sample of classical Cepheids ($>2000$) with an 
improved areal coverage was that of \citet{niko04}. The study of 
\citet{niko04} utilised multi-wavelength photometric data based on the MACHO 
(MAssive Compact Halo Objects) $V$ and $R$-band light curves 
with complete phase coverage  and single epoch photometry from 2MASS 
(Two Micron All Sky Survey) $JHK_{s}$ observations. The method developed by 
\citet{niko04} is a very robust and efficient one in the determination of 
distance and reddening to each of the individual Cepheids with respect to the 
mean distance and reddening of the LMC using the simultaneous multi-band 
photometric data. 

During the last few decades, a large number of studies have been devoted to the determination of the structure of 
the LMC using classical Cepheids. Most of the studies were based either on the 
PL or PW (period-Wesenheit) relations by employing an iterative $k\sigma$ 
clipping algorithm to clean these relations from the presence of outliers. 
Nonetheless, the sigma clipping algorithms to perform the rejection of outliers 
in these studies were arbitrary. For example, \citet{inno16} used  $6\sigma$ 
clipping before the PW relations were fitted. On the other hand, $3\sigma$ 
clipping was used by \citet{jacy16} to remove the outliers from the 
resulting PW relations. One of the important drawbacks to using such an outlier
removal algorithm is that it is based on the assumption that the errors are 
distributed normally. But it has been demonstrated by \citet{niko04} that this 
assumption is invalid for distances and reddenings.             

In the present study, we apply the method developed by \citet{niko04} to the 
simultaneously available multi-wavelength data for common classical Cepheids 
in seven photometric bands in order to independently determine the distance 
and reddening of individual Cepheids with respect to the mean distance and 
reddening of the LMC. Section~\ref{sec:data} deals with the data and sample selection procedure. 
The present sample for classical Cepheids are based on the data available in 
seven photometric bands, viz., $V,I,J,H,K, [3.6]~\mu m$ and $[4.5]~\mu m$ 
chosen from the literature. The methodology adopted in the present study to 
fit the $PL$ relations in the 
seven photometric bands is described in Section~\ref{sec:method}. The 
application of this method to obtain the $PL$ relations with and without 
corrections of distance as well as reddening values in seven photometric bands is described in Section~\ref{sec:pl}. Distance and reddening values for each of 
the individual Cepheids with respect to the mean distance and reddening of the
LMC obtained from the simultaneous solutions of apparent distance moduli are 
also discussed in that Section.
A reddening map of the LMC is presented in Section~\ref{sec:map}. The map is 
constructed from the reddening values of individual Cepheids along with the 
information on their Cartesian coordinates derived from the values of distance 
and equatorial $(\alpha,\delta)$- coordinates as provided in the catalog of 
\citet{inno16}. Determination of the geometrical and viewing angle parameters 
of the LMC obtained from the three dimensional $(x,y,z)$ distributions for all 
the Cepheids is discussed in Section~\ref{sec:structure}. 
In Section~\ref{sec:bar_disk} we separate the bar and disk Cepheids based on 
inequalities defined in the $XY$-plane and study their orientations and offsets
using the plane fitting procedure. Based on statistical tests, we also find the 
closest and farthest parts of the LMC. The summary and conclusions of this 
study are presented in Section~\ref{sec:summary}.            
\section{Data and Sample Selection}
\label{sec:data} 
We use the $3920$ classical Cepheid mean magnitudes available in the 
$V,I,J,H,Ks$ photometric bands along with the information of periods ($P$) and 
equatorial coordinates ($\alpha,\delta$)  collected by \citet{inno16}.
The mid-infrared data for magnitudes obtained in the photometric bands 
$[3.6]~\mu m$ and $[4.5]~\mu m$ taken from the SAGE 
catalog \citep{meix06} were cross matched with \citet{inno16} sample 
using the X-Match 
service of VizieR\footnote{\url{http://vizier.u-strasbg.fr/viz-bin/VizieR}} 
based on the $(\alpha,\delta)$ values within the search radius of  
$1^{\prime\prime}$. The query returned $3639$ stars which contains single 
epoch observations in the $[3.6]~\mu m$ and $[4.5]~\mu m$ photometric bands. 
Out of $3639$ stars, $3614$ stars have magnitudes in both the two photometric 
bands. Since the amplitudes of Cepheids in the mid-infrared bands are very 
small, the single epoch measurements are taken as the approximate values of 
mean magnitudes corresponding to these bands. The mid-infrared bands are 
located in the Rayleigh-Jeans tail of the blackbody spectrum. The effect of 
temperature changes are minimum in this part of the spectrum which result into 
smaller amplitudes as well as highly symmetrical nature of the light curves  of 
Cepheids \citep{chow08,free08,chow10,scow16}. It has been 
demonstrated by \citet{chow08} that the PL relations obtained from the 
mid-infrared bands agree 
quite well with each other when these relations are obtained with or without 
random phase corrections due to the small values of amplitudes in 
these photometric bands. We have chosen only those common  stars which 
have mean magnitudes in all the seven photometric bands. Their numbers turn 
out to be $3614$ out of which there are $2112$ fundamental mode (FU) Cepheids 
and $1502$ first overtone (FO) Cepheids. 

We would like to point out important differences between the study done by 
\citet{inno16} and the present study. In the study of 
\citet{inno16}, the PW relations 
for $3920$ Cepheids at optical and Near infrared were utilised for 
relative distance determinations of individual Cepheids. Relative distances 
were then converted into individual absolute distances using the mean distance 
modulus to the LMC $\mu_{0,LMC}=18.493$ mag taken from \citet{piet13}. The 
viewing angle parameters were calculated by fitting a plane of the form 
$z=f(x,y)$ to the Cartesian $(x,y,z)$ distributions obtained from the 
transformation equations connecting $(\alpha,\delta,D) \rightarrow (x,y,z)$. 
Because PW relations were used, 
the measured distances are free from interstellar extinction. 
Based on a common sample of $\sim 2600$ Cepheids with the values of apparent 
distance moduli available in six photometric bands $V,I,J,H,K_{S}$ and $w1$, a 
reddening law was fitted simultaneously to the apparent distance moduli as a 
function of inverse wavelength to determine the true distance modulus and 
reddening for individual Cepheids. A reddening map 
was also provided by \citet{inno16} based on the reddening values of these 
$2600$ Cepheids. But the values of the true distance moduli obtained from the
reddening law fit were not used by \citet{inno16} to measure the viewing angle 
parameters of the LMC. Also, the calculation of geometrical parameters 
such as the axes ratios were not attempted in that study.

On the other hand, we do not use the PW relations to 
investigate the three dimensional structure of the LMC. The geometrical and 
viewing angle parameters of the LMC are obtained using the multi-wavelength PL 
relations for a common sample of $3614$ Cepheids in seven photometric 
bands. The relative values of distance modulus as well as reddening are 
obtained from the simultaneous fitting of the seven relative apparent distance 
moduli using a reddening law as a function of inverse wavelength 
\citep{card89}. We prove that when these PL relations are corrected using the 
same set of obtained values of reddening and distance for individual Cepheids 
from the simultaneous multi-wavelength fitting, a remarkable reduction in the 
dispersions of the all PL relations is achieved. The distance moduli and 
reddening offset values of individual Cepheids are converted into their 
absolute values using the LMC mean values of these parameters taken from the 
literature. Apart from finding the viewing angle parameters of the LMC, 
determination of its geometrical parameters such as axes ratios has also been 
carried out in the present study.              
\section{Methodology}
\label{sec:method}
We follow \citet{niko04} to derive the statistical reddening and distance to 
each of the present sample of individual 
Cepheids. Availability of the values of mean magnitudes of the Cepheids in 
multi-wavelength bands allows us to disentangle the effect of reddening and 
distance which produce a scatter in the observed PL relations of 
the Cepheids. We know that the true distance modulus $(\mu_{0})$ is related to 
the observed distance modulus $(\mu_{\lambda})$ in a particular photometric 
band $(\lambda)$ as follows:
\begin{align}
\label{eq:mu}
\mu_{0}=& \mu_{\lambda}-A_{\lambda} \nonumber \\
\Rightarrow \mu_{0}= & (\overline{m}_{\lambda}-M_{\lambda})-R_{\lambda}E(B-V)\nonumber  \\
\Rightarrow \overline{m}_{\lambda}=& M_{\lambda}+\mu_{0}+R_{\lambda}E(B-V),
\end{align}                           
where $\overline{m}_{\lambda}$ and $M_{\lambda}$ denote the apparent mean 
magnitude and absolute magnitude of a Cepheid, respectively in a particular 
photometric band $(\lambda)$. $R_{\lambda}=\frac{A_{\lambda}}{E(B-V)}$ denotes 
the ratio of total to selective absorption in photometric band $\lambda$. It 
is obtained from a reddening law and is held fixed. Cepheids obey a 
PL relation given by the Leavitt law: 
\begin{align}
M_{\lambda}=&\alpha_{\lambda}\log{P}+\beta_{\lambda}+\epsilon_{\lambda}(M,T_{\rm eff},Z,\dots),
\label{eq:elambda}
\end{align}
where $\alpha_{\lambda}$ and $\beta_{\lambda}$ denote the PL coefficients. 
$\epsilon_{\lambda}$ denotes the unknown correction term related to the stellar 
parameters such as mass $(M)$, effective temperature ($T_{\rm eff}$), 
metallicity ($Z$) or other parameters due to the non-linearity of the PL 
relation in a particular photometric band ($\lambda$). Substituting the value 
of $M_{\lambda}$ into equation~(\ref{eq:mu}), we get 
\begin{align*}
\overline{m}_{\lambda}=& \alpha_{\lambda}\log{P}+\beta_{\lambda}+\mu_{0}+R_{\lambda}E(B-V)+\epsilon_{\lambda}(M,T_{\rm eff},Z,\dots) \\
\end{align*} 
Since the general form of the correction term $\epsilon_{\lambda}$ is unknown, 
we assume that it is normally distributed with zero mean and variance $\sigma^{2}_{0,\lambda}$, i.e. $\epsilon_{\lambda} \sim N(0,\sigma^{2}_{0,\lambda})$.
The unknown function $\epsilon_{\lambda}$ is responsible for the physics 
governing the dispersion of the observed PL relation in a particular 
photometric band $(\lambda)$. This can be accommodated into the zero point 
$(\beta_{\lambda})$ resulting into $\beta^{\prime}_{\lambda}$.
Writing the above equation for each individual Cepheid in a particular 
photometric band $(\lambda)$, we have
\begin{align*}
\overline{m}_{\lambda,i}=& \alpha_{\lambda}\log{P_{i}}+\beta_{\lambda}+\mu_{0,i}+R_{\lambda}E(B-V)_{i}+\epsilon_{\lambda}(M,T_{\rm eff},Z,\dots).
\end{align*}
Let $\overline{E(B-V)}_{\rm LMC}$ and $\overline{\mu}_{LMC}$ denote the average
values of the reddening and distance modulus of the LMC. Let us denote
\begin{align*}
\Delta E(B-V)_{i}=& E(B-V)_{i}-\overline{E(B-V)}_{\rm LMC}, \\
\Delta \mu_{0,i}=&\mu_{0,i}-\overline{\mu}_{LMC}. 
\end{align*}
Therefore, we have the following equation for $\overline{m}_{\lambda}$:
\begin{align}
\label{eq:mu_final}
\overline{m}_{\lambda,i}=\alpha_{\lambda}\log{P_{i}}+\beta^{\prime}_{\lambda}+\Delta \mu_{0,i}+R_{\lambda}\Delta E(B-V)_{i}.
\end{align}
Here $\beta^{\prime}_{\lambda}$ denotes the now zero point which incorporates 
$\epsilon_{\lambda}$ and the mean values $\overline{\mu}_{LMC}$ and 
$\overline{E(B-V)}_{\rm LMC}$. The values of $\overline{m}_{\lambda}$ 
corresponding to the photometric band $\lambda=\left(V,I,J,H,Ks,[3.6],[4.5]
\right)$ and 
period $(P)$ for each star are given. The central wavelengths of the seven 
photometric bands are chosen as ${\lambda_{V,I,J,H,Ks,[3.6,4.5]}}$=\{$0.555,
0.790,1.235,1.662,2.159,3.550$,\\ 
$4.493\}~\mu$m \citep{bona10}.  Using the 
\citet{card89} reddening law, the corresponding reddening values are found to 
be $R_{\lambda}=\{3.23,2.05,0.94,0.58,0.38,0.17,0.12\}$ with the ratios of 
interstellar extinction as 
$\frac{A_{\lambda}}{A_{V}}=\{1.0,0.64,0.29,0.18,0.12,0.05,0.04\}$. 
The above system of equations given by equation ~(\ref{eq:mu_final}) can be 
solved using a general linear least-squares minimisation method to find the 
PL coefficients ($\alpha_{\lambda},\beta^{\prime}_{\lambda}$) for each 
particular photometric band. This information obtained from the multi-wavelength
band PL relations can be used to obtain the values of reddening and distance 
for an individual Cepheid. Equation~(\ref{eq:mu_final}) can be solved in two 
iterations. In the first iteration, the coefficients $(\alpha_{\lambda},\beta^{\prime}_{\lambda})$ are obtained by approximating the observed mean apparent 
magnitude $\overline{m}_{\lambda}$ as:
\begin{align*}
\overline{m}_{\lambda,i} \approx \alpha_{\lambda}\log{P_{i}}+\beta^{\prime}_{\lambda}.
\end{align*}
The values of these coefficients $(\alpha_{\lambda},\beta^{\prime}_{\lambda})$
obtained in this iteration obtained from the multi-wavelength photometric band
$(\lambda)$ are then used to calculate the reddening and distance to each 
individual Cepheid. The least-squares fitting here is done using 
unweighted method with the weights $w_{i}=1.0$ and does not involve the 
magnitude uncertainty for each of the Cepheids in a particular photometric 
band. In the second iteration, equation~(\ref{eq:mu_final}) is solved  using 
the weighted least-squares minimisation method taking into account the 
distance and reddening values obtained from the first iteration with the 
weights $w_{i}=\sigma_{i}^{-2}$, where
\begin{align*}
\sigma_{i}^{2}=&\sigma_{0}^{2}+\sigma^{2}_{\rm phot}+\sigma^{2}_{\alpha}\log^{2}{P_{i}}+\sigma^{2}_{\beta}.
\end{align*}
Here $\sigma_{0}$, $\sigma_{\rm phot}$, $\sigma_{\alpha}\log{P_{i}}$ 
and $\sigma_{\beta}$ denote the  intrinsic error, photometric errors, the
errors in the slopes and intercepts of the PL relations obtained in iteration 
$1$, respectively. Following \citet{niko04}, we assume $\sigma_{0}=0.05$ mag, 
irrespective of the photometric band. 
The distance and reddening values are obtained in the 
following way after ($\alpha_{\lambda}, \beta_{\lambda}^{\prime}$) are 
determined from the first iteration:
\begin{align}
\label{eq:redd}
\Delta \mu_{\lambda,i}=& \overline{m}_{\lambda,i}-\left(\alpha_{\lambda}\log{P_{i}}+\beta_{\lambda}^{\prime} \right) \nonumber \\
\Rightarrow \Delta \mu_{\lambda,i}=&\Delta \mu_{0,i}+R_{\lambda}\Delta E(B-V)_{i}.  
\end{align}
The true distance modulus $\Delta \mu_{0,i}$ and reddening $\Delta E(B-V)_{i}$ 
for each star with respect to the mean values of the LMC can be obtained 
from iteration $1$ by plotting the apparent relative distance moduli 
$\Delta \mu_{\lambda,i}$ as a function of inverse wavelength 
$\lambda^{-1}$ for seven photometric bands with the assumption that wavelength 
dependence of these relative distance moduli is due to extinction. The amount 
of extinction at a given wavelength is inversely proportional to the 
wavelength \citep{free88}.  All the data are then fitted simultaneously with 
an interstellar extinction law using a least square fit based on $\chi^{2}$ 
minimisation whose intercept and slope yield the values of $\Delta \mu_{0,i}$ 
and $\Delta E(B-V)_{i}$, respectively along with their errors 
\citep{free88,free90,rich14,scow16}.  Substitutions of these values in 
equation~(\ref{eq:mu_final}) give the much improved PL relations with reduced 
dispersions for iteration $2$. The solutions for the present 
sample of FU and FO Cepheids are obtained separately. Once the distances are 
derived using the PL relations, they are combined together with their 
corresponding equatorial coordinates $(\alpha,\delta)$ to obtain the Cartesian 
three dimensional distribution of these Cepheids with respect to the center of 
the LMC. In order to derive the geometrical and viewing angle parameters, we 
have selected only those stars which are within $3\sigma$ limit from the mean 
of the resulting distance distributions. This ensures that the derived 
geometrical and viewing angle parameters of the LMC are unaffected by the 
distance outliers. Out of a total of $3614$ stars, $3540$ stars have 
been selected based on the $3\sigma$ limit which accounts for $98\%$ of the 
stars in the sample to find the geometrical and viewing angle parameters of 
the LMC. Of the remaining $74$ stars,  $67$ lie within $40$~kpc and 
belong to the foreground objects of the LMC. Rest of the $7$ stars are the 
background Cepheids of the LMC located at distances of more than $60$ kpc. 
Some of these outliers may be the tracers of lensing populations towards the 
LMC. 
\section{Multi-wavelength PL relations for classical Cepheids}
\label{sec:pl}
\begin{figure*}
\begin{center}
\includegraphics[width=1.0\textwidth,keepaspectratio]{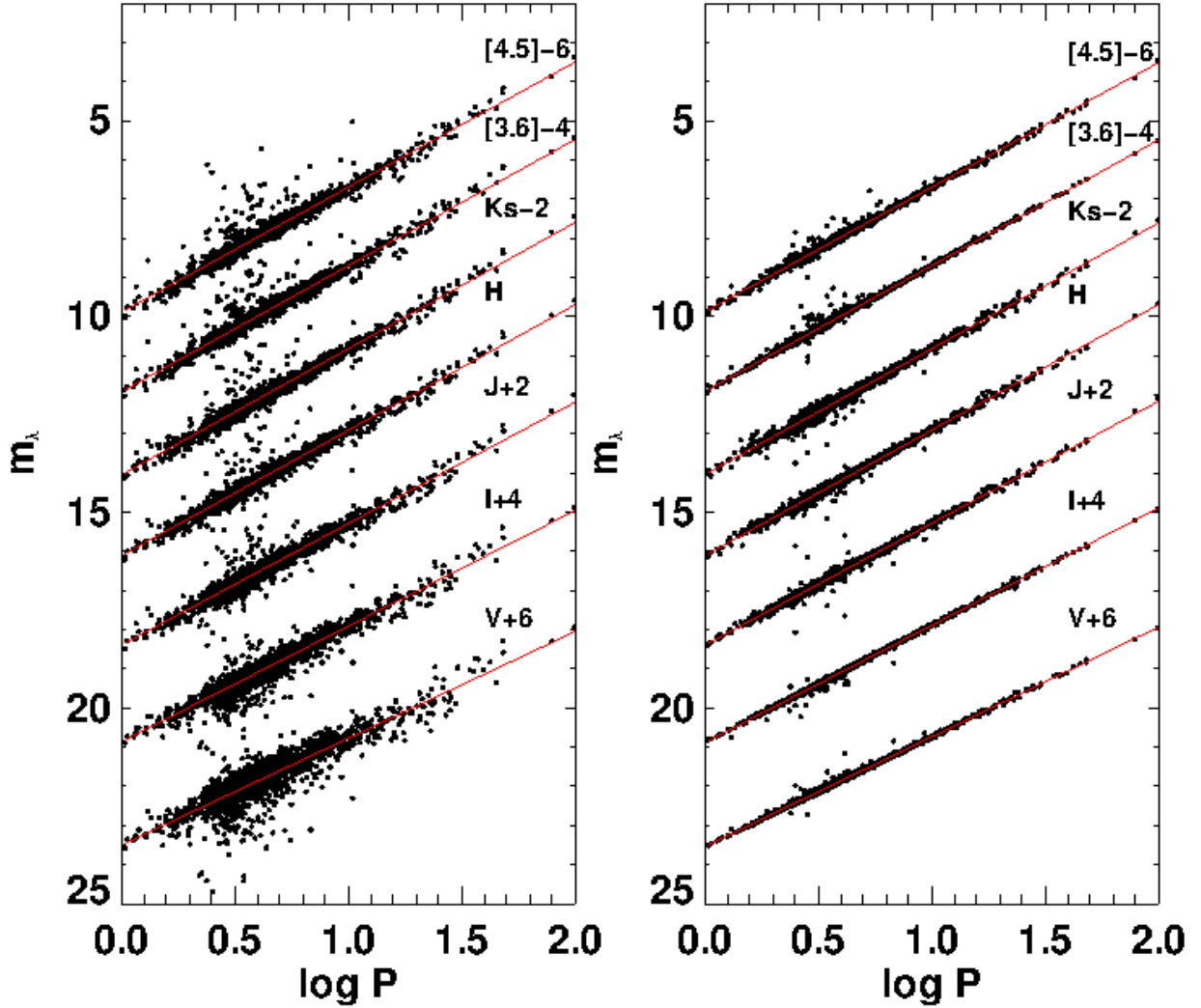}
\caption{Leavitt law for FU Cephieds.  The left panel shows the fitted PL 
relations to the observed magnitudes obtained using iteration $1$. The right 
panel shows fits to the observed magnitudes obtained in iteration $2$ 
corrected for distance and reddening found from iteration $1$. }
\label{fig:pl1}
\end{center}
\end{figure*}

\begin{figure*}
\begin{center}
\includegraphics[width=1.0\textwidth,keepaspectratio]{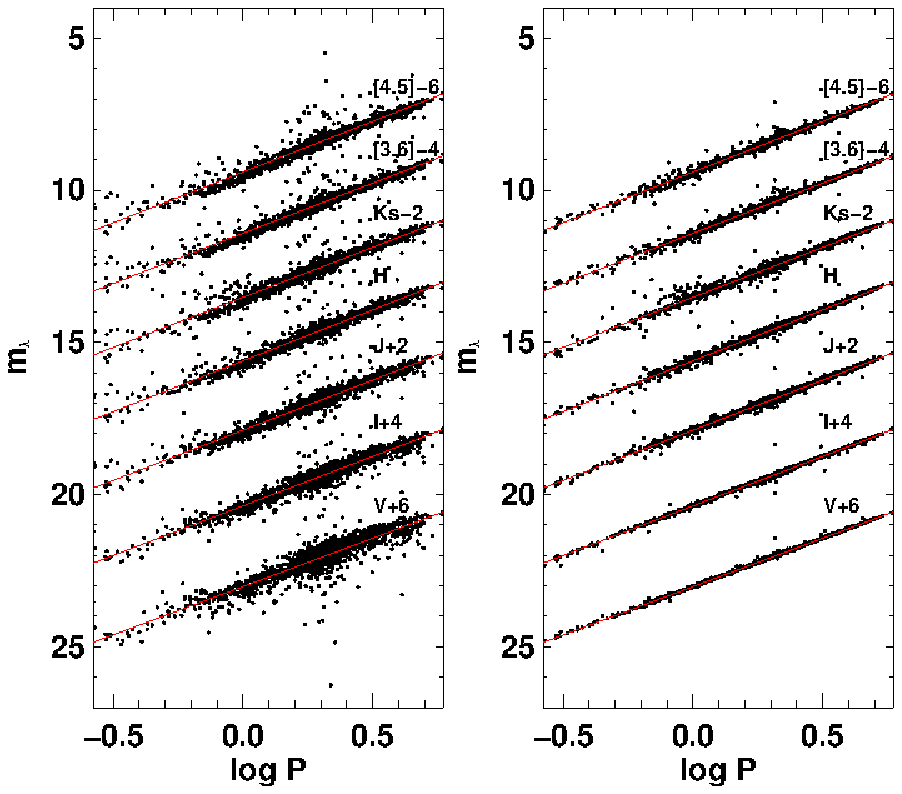}
\caption{Leavitt law for FO Cephieds. The left panel shows the fitted PL 
relations to the observed magnitudes obtained using iteration $1$. The right 
panel shows fits to the observed magnitudes obtained in iteration $2$ 
corrected for distance and reddening found from iteration $1$.}
\label{fig:pl2}
\end{center}
\end{figure*}

\begin{table*}
\begin{center}
\caption{Fitted PL relations for the LMC FU Cepheids in the multi-wavelength photometric bands $\{V,I,J,H,Ks, [3.6], [4.5]\}$. Iterations $1$ and $2$ refer to 
the PL relations without and with the observed magnitudes corrected for 
distances and reddenings, respectively.}  
\label{eq:pl1}
\begin{tabular}{ccccr} \\\hline \hline
Band & $\alpha_{\lambda}$ & $\beta_{\lambda}^{\prime}$ & $\sigma_{\rm fit}$ & $N$ \\
\hline
& &Iteration $1$& & \\ \hline 
$V.............$ & $-2.728\pm 0.095$ & $17.504\pm0.062$ & $0.312$ & $2112$ \\ 
$I.............$ & $-2.954\pm 0.095$ & $16.866\pm0.062$ & $0.218$ & $2112$ \\ 
$J.............$ & $-3.102\pm 0.095$ & $16.394\pm0.062$ & $0.179$ & $2112$ \\ 
$H............$ & $-3.203\pm 0.095$ & $16.113\pm0.062$ & $0.175$  & $2112$\\ 
$Ks............$ & $-3.226\pm 0.095$ & $16.052\pm0.062$ & $0.178$ & $2112$ \\ 
$[3.6]............$ & $-3.224\pm 0.095$ & $15.929\pm0.062$ & $0.201$ & $2112$ \\
$[4.5]............$ & $-3.195\pm 0.095$ & $15.896\pm0.062$ & $0.196$ & $2112$ \\
\hline
& & Iteration $2$ && \\ \hline 
$V.............$ & $-2.728\pm 0.003$ & $17.504\pm0.002$ & $0.031$ & $2112$  \\ 
$I.............$ & $-2.954\pm 0.003$ & $16.865\pm0.002$ & $0.035$ & $2112$ \\ 
$J.............$ & $-3.101\pm 0.005$ & $16.392\pm0.004$ & $0.057$ & $2112$ \\ 
$H............$ & $-3.202\pm 0.006$ & $16.111\pm0.004$ & $0.066$ & $2112$ \\ 
$Ks............$ & $-3.224\pm 0.008$ & $16.051\pm0.005$ & $0.081$& $2112$  \\
$[3.6]............$ & $-3.222\pm 0.007$ & $15.927\pm0.004$ & $0.070$ & $2112$ \\
$[4.5]............$ & $-3.194\pm 0.006$ & $15.894\pm0.004$ & $0.066$ & $2112$ \\\hline
\end{tabular}
\end{center}
\end{table*}
\begin{table*}
\begin{center}
\caption{Fitted PL relations for the LMC FO Cepheids in the multi-wavelength photometric bands $\{V,I,J,H,Ks, [3.6],[4.5]\}$. Iterations $1$ and $2$ refer 
to the PL relations without and with the observed magnitudes corrected for 
distances and reddenings, respectively.}
\label{eq:pl2}
 \begin{tabular}{ccccr} \\\hline \hline
Band & $\alpha_{\lambda}$ & $\beta_{\lambda}^{\prime}$ & $\sigma_{\rm fit}$ & $N$ \\
\hline
& &Iteration $1$& & \\ \hline 
$V.............$ & $-3.171\pm 0.124$ & $17.022\pm0.044$ & $0.339$ & $1502$  \\ 
$I.............$ & $-3.252\pm 0.124$ & $16.372\pm0.044$ & $0.231$ & $1502$ \\
$J.............$ & $-3.294\pm 0.124$ & $15.885\pm0.044$ & $0.193$ & $1502$ \\ 
$H............$ & $-3.342\pm 0.124$ & $15.604\pm0.044$ & $0.187$  & $1502$\\ 
$Ks............$ & $-3.284\pm 0.124$ & $15.520\pm0.044$ & $0.219$ & $1502$ \\ 
$[3.6]............$ & $-3.295\pm 0.124$ & $15.416\pm0.044$ & $0.196$&$1502$ \\
$[4.5]............$ & $-3.329\pm 0.124$ & $15.404\pm0.044$ & $0.203$ & $1502$ \\\hline
& & Iteration $2$ && \\ \hline 
$V.............$ & $-3.172\pm 0.004$ & $17.022\pm0.001$ & $0.029$ & $1502$  \\ 
$I.............$ & $-3.251\pm 0.004$ & $16.372\pm0.002$ & $0.037$ & $1502$ \\ 
$J.............$ & $-3.292\pm 0.010$ & $15.884\pm0.004$ & $0.079$ & $1502$ \\ 
$H............$ & $-3.339\pm 0.010$ & $15.603\pm0.004$ & $0.084$ & $1502$ \\ 
$Ks............$ & $-3.280\pm 0.015$ & $15.519\pm0.005$ & $0.120$& $1502$  \\ 
$[3.6]............$ & $-3.292\pm 0.010$ & $15.415\pm0.003$ & $0.077$&$1502$ \\
$[4.5]............$ & $-3.326\pm 0.011$ & $15.403\pm0.004$ & $0.086$ & $1502$ \\
\hline
\end{tabular}
\end{center}
\end{table*}
The results of PL relations obtained using the methodology as described in 
Section~\ref{sec:method} are listed in Tables~\ref{eq:pl1} and \ref{eq:pl2}
for FU and FO Cepheids, respectively. Both the iterations, with and without 
distance/reddening correction are listed. 
The resulting plots thus obtained from the analysis are shown in Figs.~\ref{fig:pl1} and \ref{fig:pl2}. The observed dispersions of the PL relations as in 
iteration 1 are due to the four factors: (1) intrinsic variation due to the 
$\epsilon_{\lambda}$ term in equation~(\ref{eq:elambda}), (2) distance (3) 
reddening and (4) photometric errors. In iteration 2 of the fitted PL 
relations, distance and reddening are corrected and the observed dispersions 
get smaller after the corrections which account for the intrinsic variation 
due to the $\epsilon_{\lambda}$ terms as well as photometric errors in the 
observed magnitudes in the given photometric bands. It should be noted from 
the plots of the in Figs.~\ref{fig:pl1} and \ref{fig:pl2} that the dispersions 
of the observed magnitudes in the left panels are more for the optical and 
decrease as we go towards longer wavelengths (See third columns of 
Tables~\ref{eq:pl1} and \ref{eq:pl2}). This is because interstellar 
extinction is more in the optical and decreases as one goes towards near and 
mid-infrared bands. However, from the right panels
of Figs.~\ref{fig:pl1} and \ref{fig:pl2}, we see the contrast behaviour 
where the observed magnitudes are corrected for distances and interstellar
extinction. The dispersions increase as we go from the optical to the near 
and mid-infrared in which the observed magnitudes are corrected for distances 
and interstellar extinction. This is due to the fact that as the 
wavelength increases, the amplitudes of the Cepheid light curves decrease and 
hence it becomes increasingly difficult to derive their mean magnitudes with  
precision.

The PL relations corrected for extinction using the \citet{zari04} 
extinction map  obtained for the LMC FU Cepheids derived by \citet{chow09} 
based on the OGLE-III data are consistent with the relations found in the 
present study. We also find that the respective PL relations obtained by 
\citet{inno16} using the reddening correction determined from the simultaneous 
multi-wavelength fitting of apparent distance modulus as well as making use of 
the \citet{hasc11} extinction map for the LMC are consistent with the relations 
derived here within the quoted uncertainties. However, the number of stars 
varied from $1112$ to $1526$ while deriving the PL relations in the study of 
\citet{inno16}. Unlike the use of different number of FU Cepheids as in the 
study of PL relations by \citet{inno16}, we have used the same number of stars 
($2112$) in all the seven photometric bands to derive the PL relations in 
their respective bands. The dispersions in all the PL relations reduce 
significantly when corrected for distance and reddening values obtained from 
the simultaneous solutions of apparent distance moduli calculated in 
seven photometric bands for individual Cepheids. This reduction in the 
dispersion for many of the PL relations discussed here is remarkably more 
than those obtained by \citet{inno16}. For instance, the reduction in the 
dispersion of the PL relations for the highly accurate and precise optical 
OGLE-IV V- and I-band data is more than a factor of two as compared to those 
obtained by \citet{inno16}. This fact proves that the reddening values 
determined in the present study are robust and supports the evidence of a 
highly accurate reddening map of the LMC. However, because of increasing 
uncertainties associated with the determinations of the near-infrared mean 
magnitudes, the dispersions turn out to be of the same order of magnitudes in 
the present study as those obtained by \citet{inno16} in these bands.   

One of the important results of the analysis of the
present sample of FU Cepheids is the mid-infrared PL relations at $[3.6]~\mu m$ 
and $[4.5]~\mu m$ bands. The slopes of the two relations obtained in this 
study $-3.222\pm 0.007$ and $-3.194\pm 0.006$ are comparable to the slopes of 
$-3.253\pm 0.010$ and $-3.214\pm 0.010$ obtained for the two bands by 
\citet{chow09} utilising more than $1600$ FU Cepheids based on the IRAC-band 
data taken from the SAGE Winter'08 Archive. 
On the other hand, based on the argument that the mid-infrared luminosity lies
at the Rayleigh-Jeans tail of the blackbody function, the predicted slopes of 
the PL relations at the mid-infrared will be $\sim -3.23$ \citep{neil10} which 
is quite close to the slopes for the two mid-infrared bands used in the present 
study. The slope of $-3.194\pm 0.007$ obtained by \citet{inno16} with the mean 
magnitudes determined from the ALLWISE multi-epoch catalog is also in good 
agreement with the slope  $-3.222\pm0.007$ for $[3.6]~\mu m$ PL relation 
obtained from the single-epoch SAGE catalog in the present study. 

For the LMC FO Cepheids, it can be seen that the PL relations obtained by 
\citet{inno16} for six multi-wavelength photometric bands when corrected 
for extinction using the reddening values determined from the reddening law fitting method for individual FO Cepheids are drastically different than 
those obtained using the \citet{hasc11} reddening map (For difference between 
the two PL relations obtained for the same photometric band, see Table~8 
of \citet{inno16} for FO Cepheids). For example, the PL slopes in the $V$-band 
when extinction corrections are applied using the reddening values 
obtained from these two different means are $-3.080\pm 0.003$ and 
$-3.299\pm 0.004$, respectively. These two values are significantly different 
from each other at more than $6\sigma$ level. Such huge differences in the 
PL slopes for other photometric bands are also quite noticeable when 
extinction corrections were 
applied using these two different means of reddening estimations in the study 
of \citet{inno16}. It is also quite interesting to see that the slope of the PL
relation for FO Cepheids in the $V$-band, $-3.299$ obtained from the 
extinction correction using the \citet{hasc11} reddening map is exactly the 
same as that obtained \citet{bhar16} which has also been obtained using the 
same reddening map of \citet{hasc11}. Nonetheless, this discrepancy raises a 
serious concern regarding the use of reddening values obtained from various 
sources while deriving the extinction corrected PL relations for FO Cepheids.

There are two possible reasons why the slopes of the extinction corrected PL 
relations differ by large values. Firstly, due to the use of different numbers 
of FO Cepheids by \citet{inno16} to derive the PL relations while corrected for 
extinction using these two sources of reddening estimations. The number of 
Cepheids used  to derive the extinction-corrected $V$-band PL relation by
\citet{inno16} using the reddening values obtained from the reddening law 
fitting method and the Haschke reddening map were $1056$ and $795$ in number 
with the dispersions of $\sigma=0.09$ and $0.19$, respectively. Secondly, due 
to the different levels of uncertainties in the reddening values obtained from 
these two sources. Comparison with the $V$-band PL slope from \citet{bhar16} 
rules out the first option, in which the same value as that of \citet{inno16} 
having a dispersion of $\sim 0.18$ was obtained using $1084$ 
number of Cepheids where the reddening correction was done with the help of 
the Haschke reddening map. Hence we are left only with the second option, 
i.e., the significant steeper slopes of the PL relations obtained from the 
Haschke reddening map than those obtained from the reddening law fitting 
method by \citet{inno16} or in the present study are due to large 
uncertainties in the reddening values of individual stars obtained from the 
Haschke reddening map. In fact, for many of the stars we find that the errors 
in the reddening values obtained from the \citet{hasc11} reddening map are of 
the order of magnitude of the reddening values themselves or sometimes more. 
For an instance, the average value of the reddening in the \citet{hasc11} 
reddening map is $\sim 0.09$ mag, whereas the average value of the reddening 
errors is $\sim 0.10$ mag. Therefore, for the individual Cepheids that have 
dominating reddening uncertainties, their dereddened magnitudes will have large 
uncertainties. Nonetheless, we find that the \citet{hasc11} reddening map is 
quite consistent with the reddening map obtained in the present study. 

It should be noted that reddening causes more 
dispersions of the PL relations in the optical bands and less to the infrared 
and mid-infrared bands. From the PL relations obtained for the FO Cepheids as 
given in Table~\ref{fig:pl2}, one can find that the dispersion of the $V$-band 
PL relation reduces from $0.339$ to $0.029$ when the extinction corrections are 
applied using the reddening values obtained from the fitting of the observed 
apparent distance moduli in seven multi-wavelength photometric bands as a 
function of inverse wavelength in the present study. This demonstrates the 
fact that accurate values of reddening estimations will cause the dispersions 
of all the PL relations to reduce significantly while inaccurate values will 
make a slight reduction in the dispersions and may lead to  systematic 
biases in the PL slopes. The \citet{hasc11} reddening map was constructed from 
the Red Clump stars which are of intermediate-age (ages $\approx 2-9$ Gyr) as 
compared to the young stellar populations such as classical Cepheids 
(ages $\approx 10-300$ Myr) used in the present study. Just as it has been 
seen that applying extinction maps from young OB stars in the LMC to older Red 
Clump stars in the same regions yields incorrect distances by $0.2$ mag, it 
may also be inaccurate to find reddening values on a star-by-star basis for 
individual Cepheids using reddenings derived from Red Clump stars. Dust 
affects different stellar populations differently \citep{zari99,subr05}.
Besides the multi-wavelength 
photometric data available for Cepheids in more than one band as in the 
present study is very much suitable to directly determine the reddening values 
for individual stars rather than relying on various extinction maps. This is 
particularly useful when the targets are affected due to the high foreground 
and spatially variable reddening.                      
\section{Reddening map of the LMC}
\label{sec:map}
\begin{figure*}
\begin{center}
\includegraphics[width=1.0\textwidth,keepaspectratio]{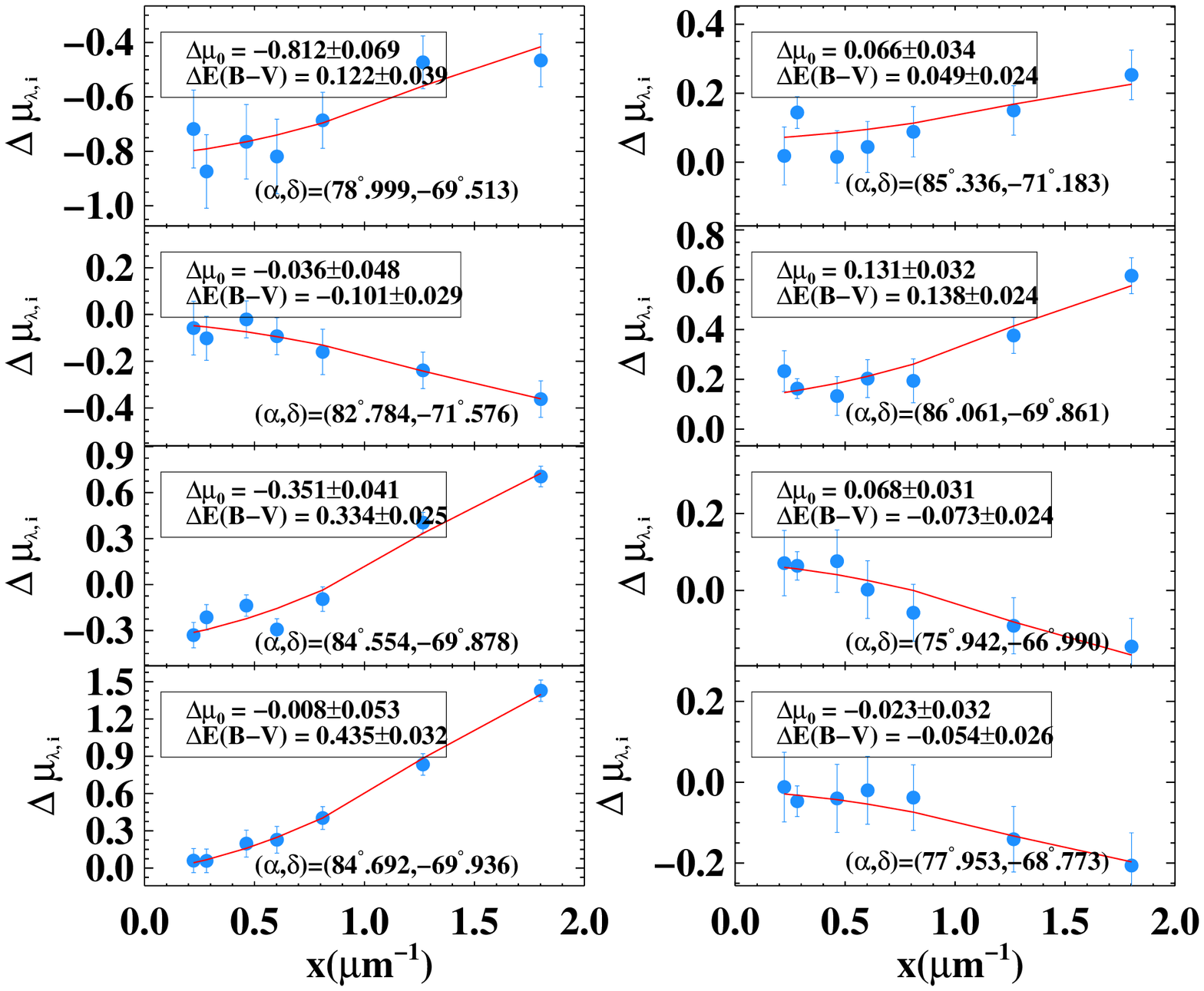}
\caption{The method of calculation of true distance modulus $\Delta \mu_{0,i}$
and reddening $\Delta E(B-V)_{i}$. Each of the data points corresponds to  the
distance modulus $\Delta \mu_{\lambda,i}$  determined for the seven 
photometric bands plotted as a function of inverse of their respective 
wavelengths $(\lambda^{-1})$. A reddening law \citep{card89} is fitted 
simultaneously to the seven distance moduli using the PL relations obtained
from iteration~1 as listed in Tables~\ref{fig:pl1} anf \ref{fig:pl2}  based on 
the $\chi^{2}$ minimisation. The solid line shows the best fit
reddening law which yields the extinction corrected distance modulus $\Delta \mu_{0,i}$ and reddening $\Delta E(B-V)_{i}$.} 
\label{fig:redd_fit}
\end{center}
\end{figure*}

\begin{figure}
\begin{center}
\includegraphics[width=0.5\textwidth,keepaspectratio]{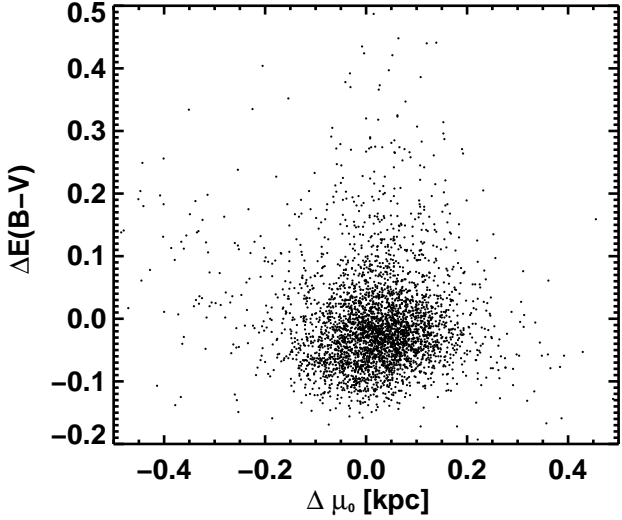}
\end{center}
\caption{$\Delta E(B-V)$ vs $\Delta \mu_{0}$ for all Cepheids in the present 
sample (FU and FO). These values are found to uncorrelated which reflects the
independent and unbiased determination of these two quantities using the 
methodology as described in Section~\ref{sec:method}.}
\label{fig:ebv_mu0}
\end{figure}

\begin{figure}
\begin{center}
\includegraphics[width=0.5\textwidth,keepaspectratio]{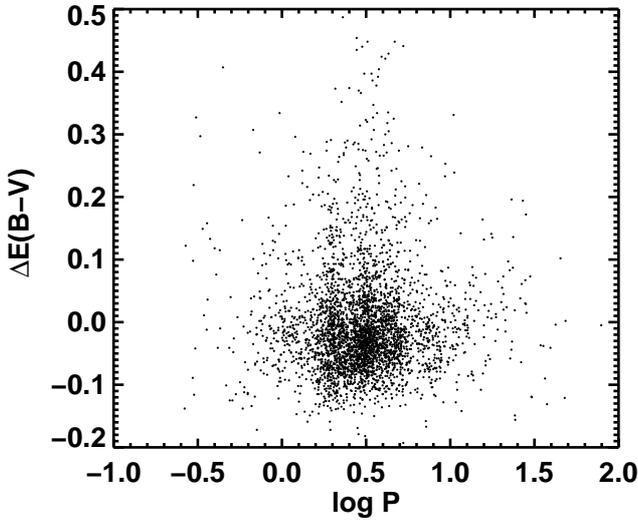}
\caption{Reddening values as a function of the logarithm of period. The plot 
suggests that there is no systematic error due to the given PL relations.}
\label{fig:pebv}
\end{center}
\end{figure}

The values of the true distance modulus $\Delta \mu_{0,i}$ and 
$\Delta E(B-V)_{i}$ for a representative sample of eight stars in the present
analysis  is shown in Fig.~\ref{fig:redd_fit}. The mutiband apparent distance 
moduli for seven photometric bands as a function of inverse wavelength 
$x=\lambda^{-1}$ are plotted. A reddening law \citep{card89} is then fitted 
simultaneously to the data which yields the extinction corrected distance modulus $\Delta \mu_{0,i}$ and reddening $\Delta E(B-V)_{i}$ for a particular
star. A plot of $\Delta E(B-V)$ vs $\Delta \mu_{0}$ for all Cepheids in 
the present sample (FU and FO) is shown in Fig.~\ref{fig:ebv_mu0}. The plot
shows that these two values are uncorrelated which reflects an independent and
unbiased determination of these two quantities using the methodology as 
described in Section~\ref{sec:method}. Fig.~\ref{fig:pebv} shows the 
reddening values as a function of the period which addresses the fact that there
is no systematic effect on the derived reddening values using the PL 
relations. The values of distance modulus $\Delta \mu_{i}$ and reddening 
$\Delta E(B-V)_{i}$ derived from equation~(\ref{eq:redd}) are offsets from the 
mean distance modulus and reddening of the LMC. In the present case, these 
values are arbitrary and cannot be determined based on the data alone.
Since the purpose of the present paper is not to calibrate the mean distance 
and reddening of the LMC, but to determine its geometrical and structural  
parameters, using the mean values of these parameters does not have any 
impact on the results \citep{niko04}. Following \citet{niko04}, we assume
$\overline{E(B-V)}_{\rm LMC}=0.14\pm 0.02$ mag and the mean value of the LMC 
distance modulus is taken as $\overline{\mu}_{\rm LMC} =18.493\pm0.002(\rm stat.)\pm 0.047(\rm sys.)$ mag \citep{piet13}. Therefore, the absolute values 
of reddening and distance modulus  of the present sample of classical Cepheids 
are obtained  from $E(B-V)_{i}=\Delta E(B-V)_{i}+0.14$ mag, 
$\mu_{0,i}=\Delta \mu_{0,i}+18.493$ mag. The use of the value of $\overline{E(B-V)}_{\rm LMC}=0.14\pm 0.02$ mag makes most of the stars in our sample to have 
$E(B-V)_{i} > 0$. For $25$ stars, we get $E(B-V) < 0$ which constitute less 
than $1\%$ of the present sample of classical Cepheids. 

\begin{figure*}
\begin{center}
\vspace{0.02\linewidth}
\begin{tabular}{ccc}
\vspace{+0.01\linewidth}
  \resizebox{0.32\linewidth}{!}{\includegraphics*{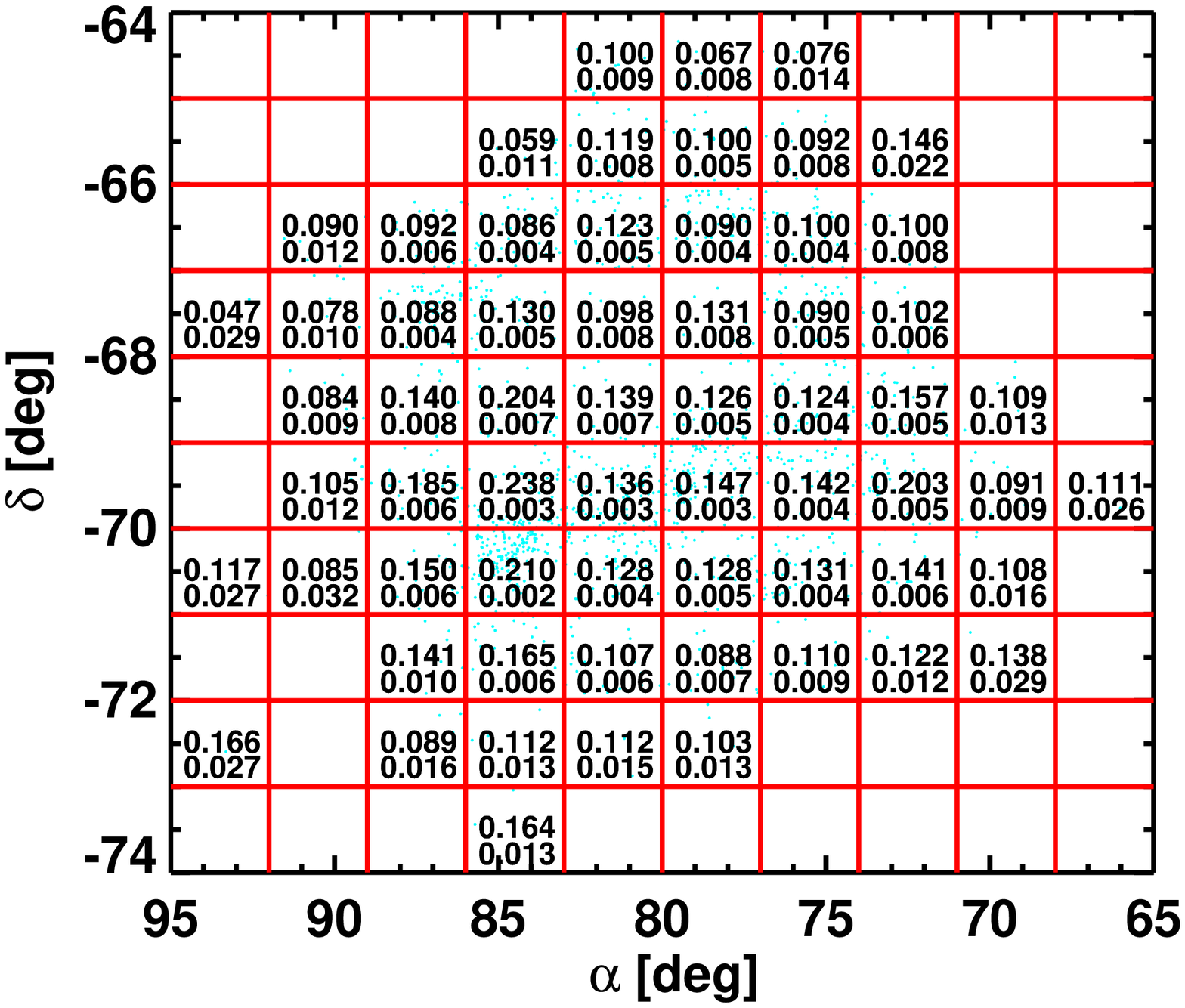}}&
\resizebox{0.32\linewidth}{!}{\includegraphics*{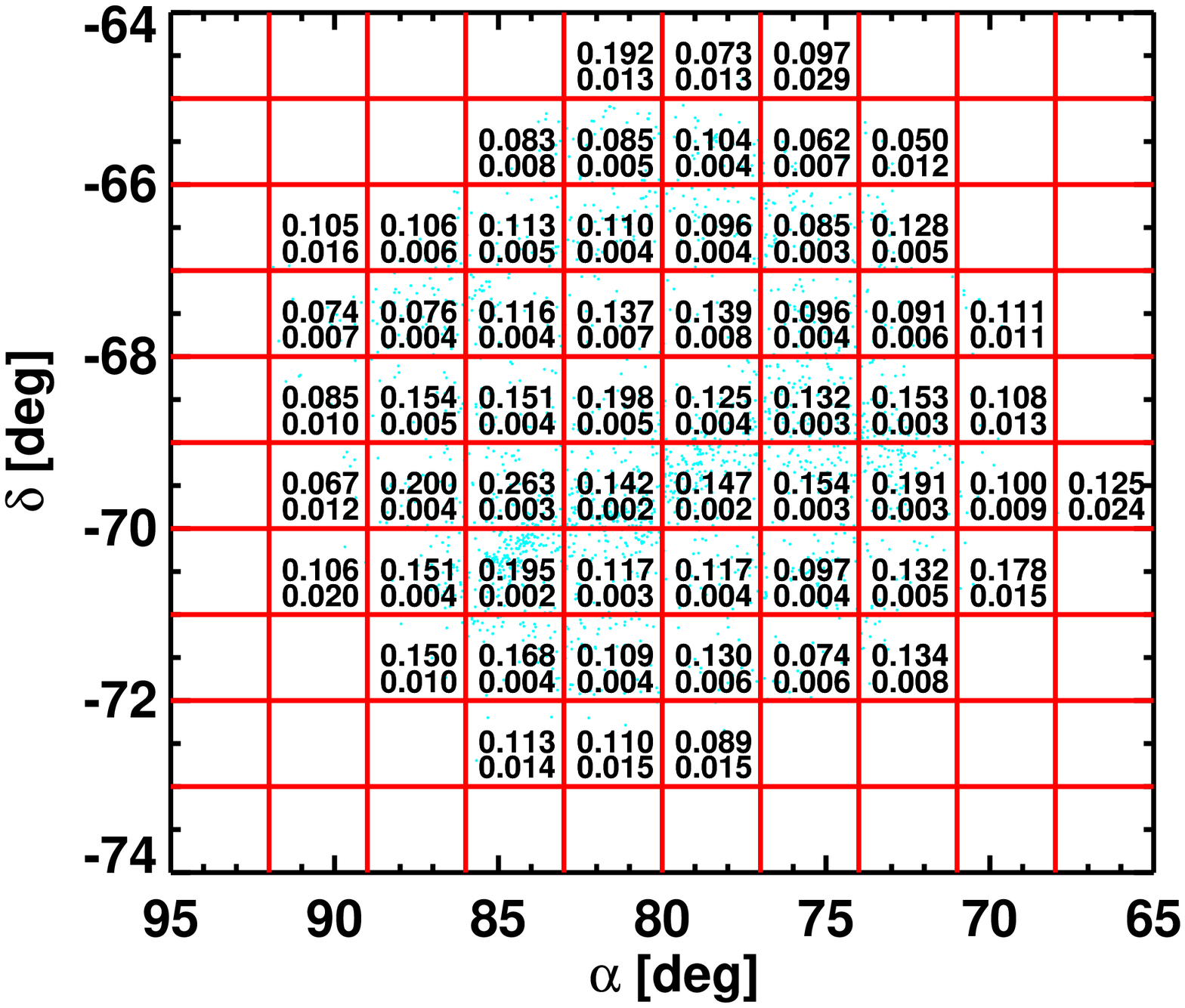}}&
  \resizebox{0.32\linewidth}{!}{\includegraphics*{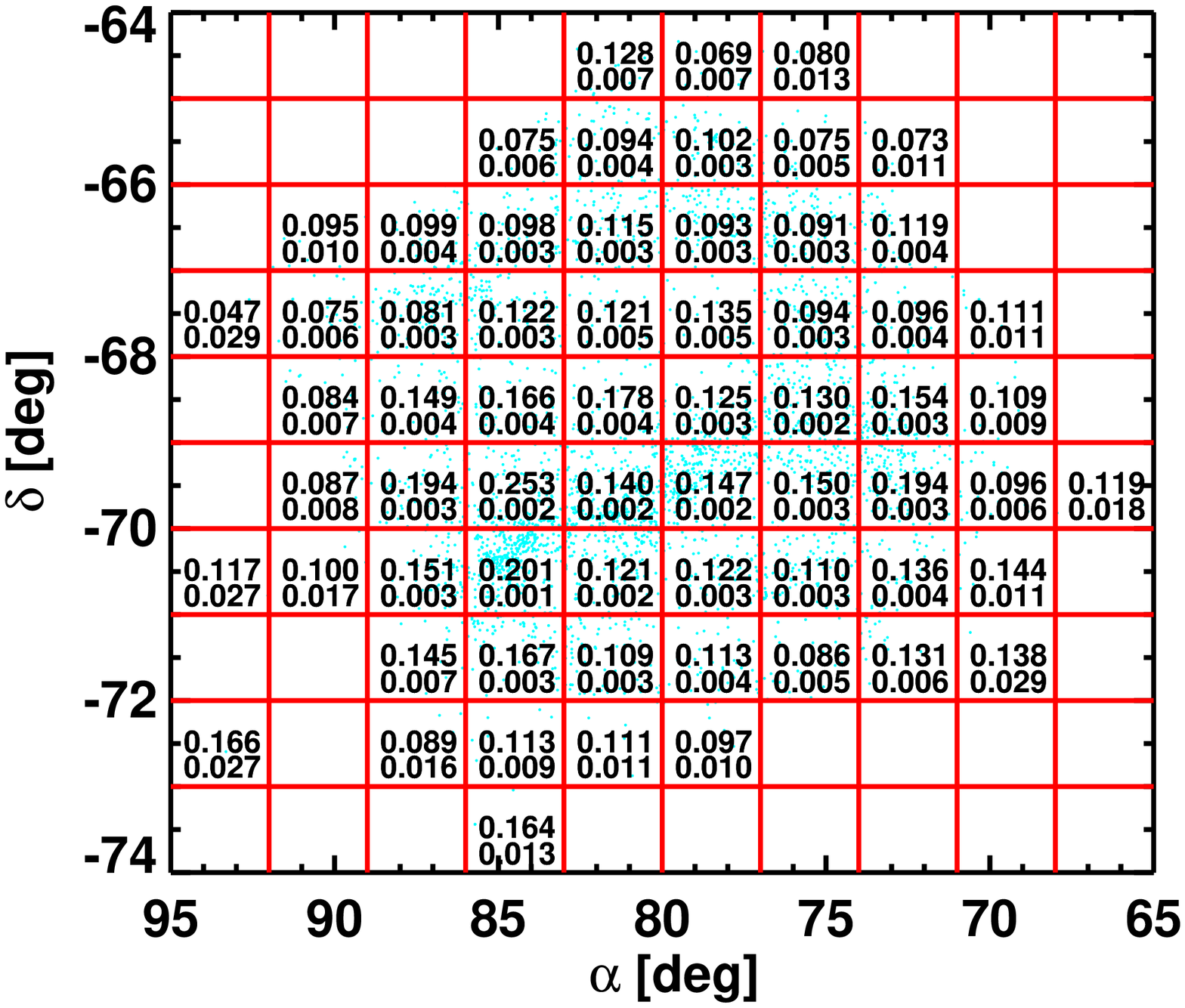}}\\
\vspace{-0.04\linewidth}
\end{tabular}
\caption{Reddening distribution $E(B-V)$ of the selected sample of FO, FU and 
(FO+FU) Cepheids in the LMC. $E(B-V)$ values are binned on a 
$10\times 10$ coordinate grid. Weighted average reddening values and their 
associated errors calculated are shown in each grid box.}
\label{reddening_dens}
\end{center}
\end{figure*}

\begin{figure*}
\vspace{0.02\linewidth}
\begin{tabular}{ccc}
\vspace{+0.01\linewidth}
  \resizebox{0.32\linewidth}{!}{\includegraphics*{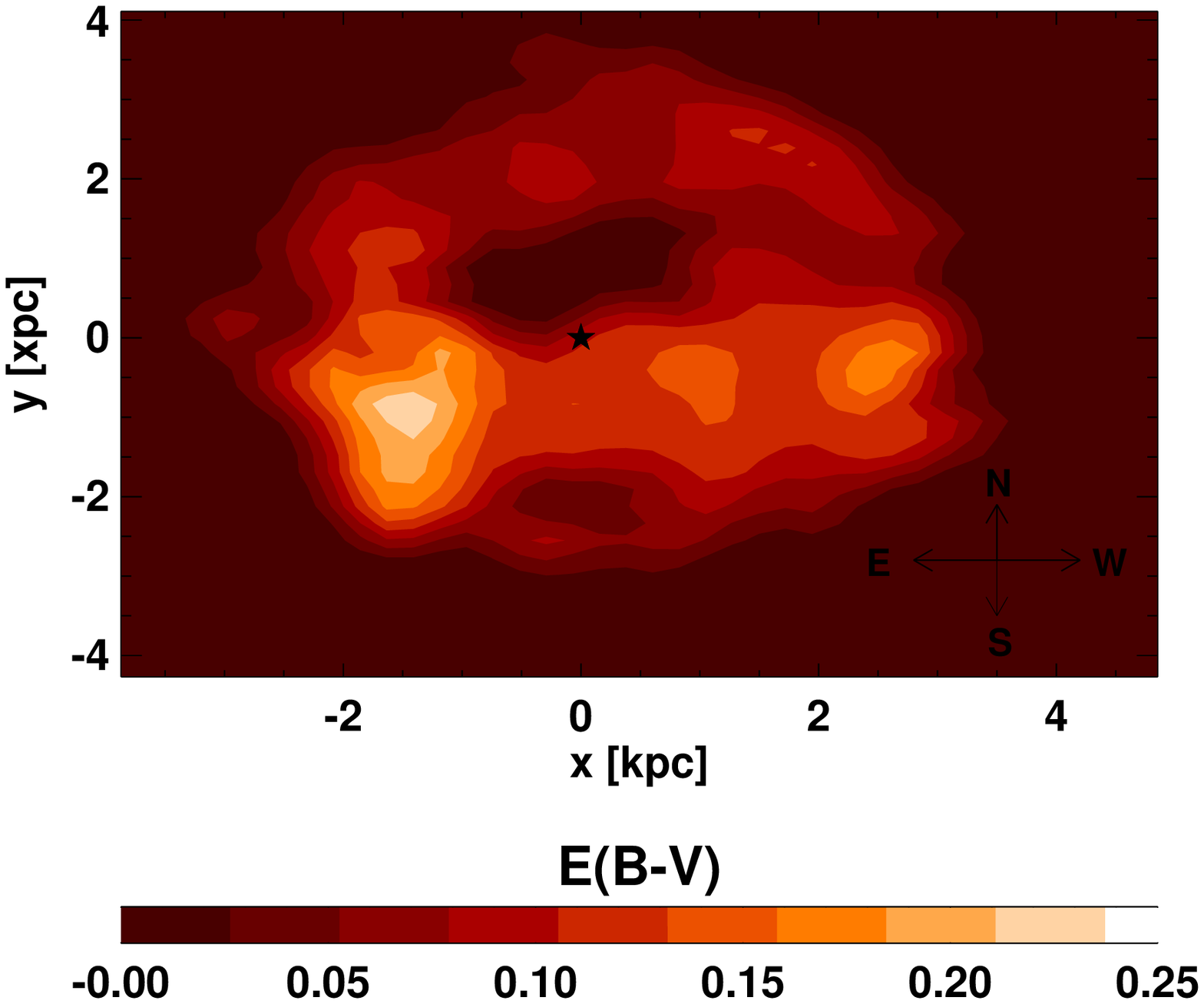}}&
\resizebox{0.32\linewidth}{!}{\includegraphics*{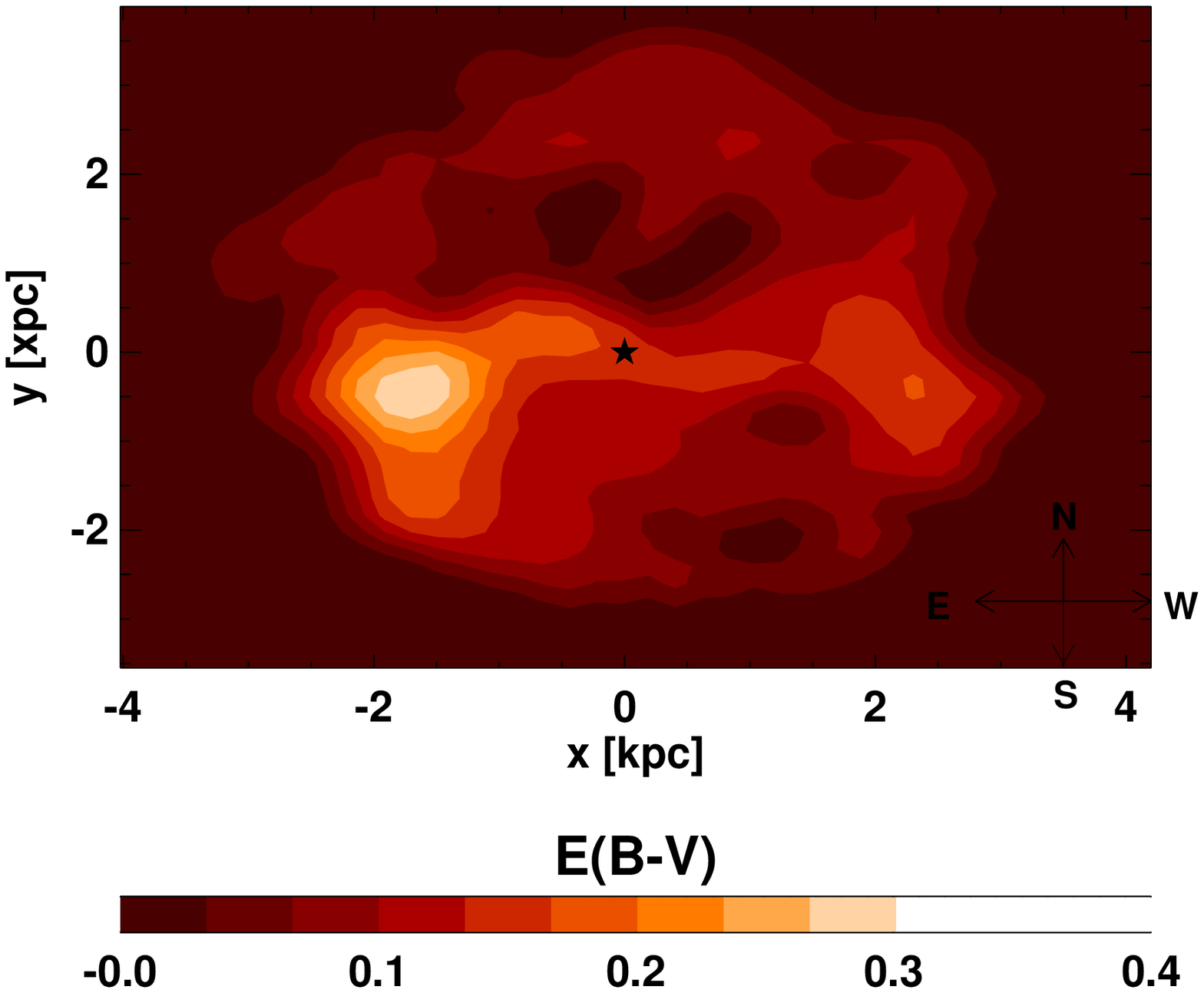}}&
  \resizebox{0.32\linewidth}{!}{\includegraphics*{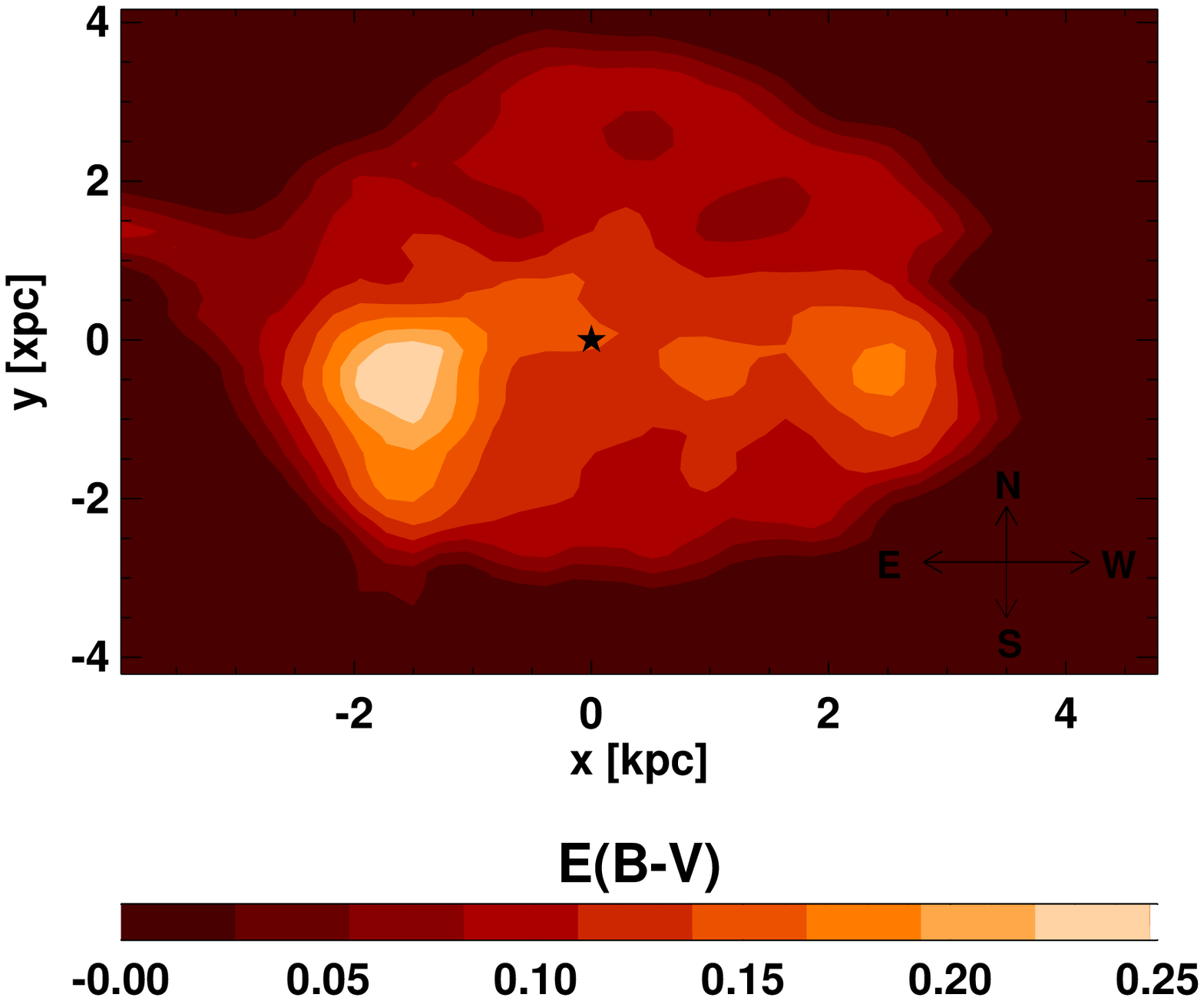}}\\
\vspace{-0.04\linewidth}
\end{tabular}
\caption{Density maps of reddening using FO, FU and combined (FO+FU) Cepheids 
are shown from left to right. The map is produced computing the average 
reddening on a $40\times 40$ grid in $(x,y)$ coordinates and smoothing the 
resulting distribution using the IDL routine \small{filter\_image}. In all the 
panels a high reddening zone near $30$ dor ($x \sim -1.8,~y \sim -1$) is quite 
easily discernible.}
\label{fig:density_map}
\end{figure*}

Once the absolute reddening values for all the stars are obtained, we proceed
towards constructing the reddening map of the LMC. Using the information of 
$(\alpha,\delta)$ for all the Cepheids as provided in the \citet{inno16} 
catalog, we first bin the observed area of the LMC on a $(10\times 10)$ 
coordinate grid. Depending on the $(\alpha,\delta)$ values, each grid will 
contain a finite number of stars with their absolute reddening values. 
Reddening distribution of the present sample of LMC classical Cepheids is 
shown in Fig.~\ref{reddening_dens}. The weighted average 
values of reddening of all the stars falling in a bin is taken to be the value 
of reddening corresponding to that bin with standard deviation as its 
statistical uncertainty.  From the reddening map, we can see that the 
reddening values in the region between $\alpha=83^{\circ}-88^{\circ}$, 
$\delta=-71^{\circ}$ to $-68^{\circ}$ have the highest values with the most 
prominent one roughly located at $\alpha \sim 85^{\circ}$ and 
$\delta=-69^{\circ}$. This region was also identified as having the highest 
reddening in the study of \citet{niko04} as well as \citet{inno16} and is 
associated with the $30$ Doradus region (Tarantula Nebula), the most active 
star forming brightest HII region in the Local group located in the bar of the 
LMC \citep{tatt13}. The 30 Doradus region is located at the center of the 
compact massive cluster RMC 136a \citep{evan11,tatt13} and contains the highest 
concentration of young clusters \citep{glat10}. Another region located in the 
vicinity of $\alpha=74^{\circ}$ and $\delta=-69^{\circ}$ has higher values of 
reddening. We identify this region to be associated with the LMC HI 
supergiant shells SGS 12 (LMC3) which hosts young star clusters 
\citep{glat10}. For better visualisation of the reddening structure, the 
density maps of reddening for FO, FU and combined (FO+FU) Cepheids are shown in 
Fig.~\ref{fig:density_map}. The maps are produced using the IDL routine  
{\small{filter\_image}} by applying $(3\times3)$ median and then $(3\times3)$ 
moving average, both applied to all the pixels. The bins with 
number of stars $\ge 3$ are used in order to reduce the noise in the 
reddening density map. The reddening map for the LMC 
obtained by \citet{niko04} and \citet{inno16} using the classical Cepheids are 
 in good agreement with the map found in the present study, although 
the number of Cepheids as well as the number of wavelength bands were less in 
earlier studies. The map is also consistent with the \citet{hasc11} 
reddening map obtained using the Red Clump stars.

We have also quantified how the reddening and distance values change if 
there were no mid-infrared photometry. The mean and standard deviation of the 
difference in the two values of $\Delta E(B-V)$ obtained with and without 
the mid-infrared photometry are $0.00$ and $0.03$ mags, respectively, whereas 
the corresponding values for distance are $0.01$ kpc and $1.68$ kpc, 
respectively.  Therefore, the individual reddening and distance values 
can differ by $0.03$ mag  and $1.68$ kpc, respectively at the $1\sigma$ 
uncertainty level. However, the means of these values in both the two cases are 
almost identical.      

\section{Three Dimensional Structure of the LMC}
\label{sec:structure}
A common objective in studying the three dimensional structure of a galaxy is to
find its angular orientation and the axes ratios in three different directions.
The orientation of the galaxy is measured by means of two angles, the 
inclination angle ($i$), and the position angle of the line of nodes
($\theta_{lon}$). The angle $\theta_{lon}$ is defined as the angle between the 
intersection of the plane of the distribution of the stars in the galaxy and 
the sky plane defined in Cartesian coordinates. Position angle in astronomy is 
measured from north ($0^{\circ}$) towards east $90^{\circ}$. The inclination angle $(i)$ is defined as the angle how the galaxy 
plane $(x^{\prime},y^{\prime})$ is inclined with respect to the sky plane 
$(x,y)$ \citep{vand01}. The combination of these two 
angles gives a full description of the orientation of a galaxy, and 
combined with its ($\alpha,\delta$) values and distance ($D$), a full 
description of its position in three dimensional space can be obtained. 
The orientation of a galaxy is an important parameter in galaxy interaction
simulations. The most commonly used method to obtain the angles $i$ and 
$\theta_{lon}$ is by fitting a plane of the form $z=f(x,y)$ to the 
distribution of the stars in Cartesian coordinates $(x,y,z)$, and 
finding these angles between that plane and the sky plane. To a reasonable 
limit, the geometry of the LMC can be considered to be planar \citep{vand01}.  
Another method, which has recently come into use, is to fit a triaxial 
ellipsoid to the Cartesian distribution of the stars, by means of a principal 
axis transformation. Apart from finding these two angles, this method also 
yields the values of the axes ratios of the galaxy in three perpendicular 
directions \citep{paz06,pejc09,subr12,deb14}. 
          
\begin{figure*}
\vspace{0.02\linewidth}
\begin{tabular}{ccc}
\vspace{+0.01\linewidth}
  \resizebox{0.32\linewidth}{!}{\includegraphics*{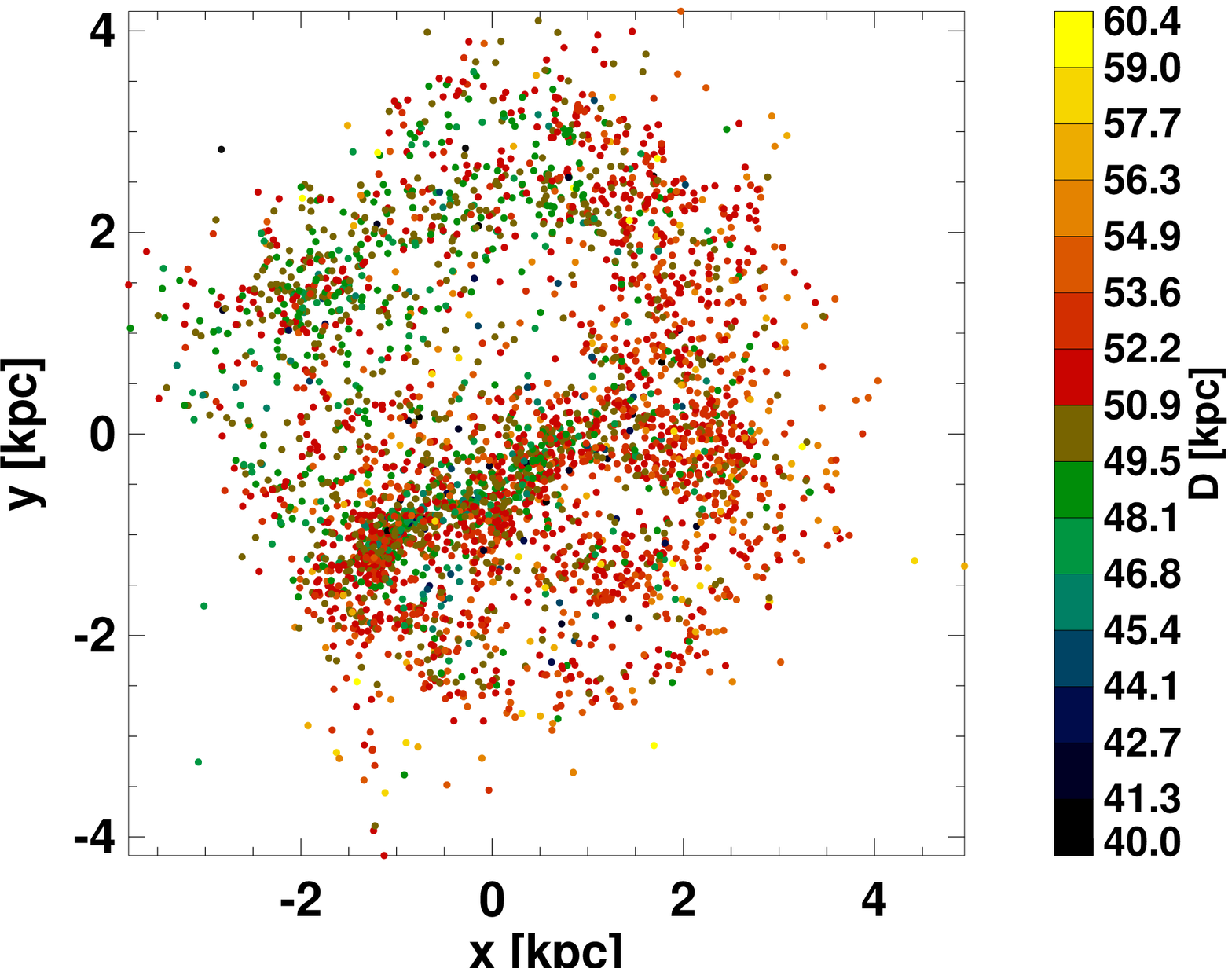}}&
\resizebox{0.32\linewidth}{!}{\includegraphics*{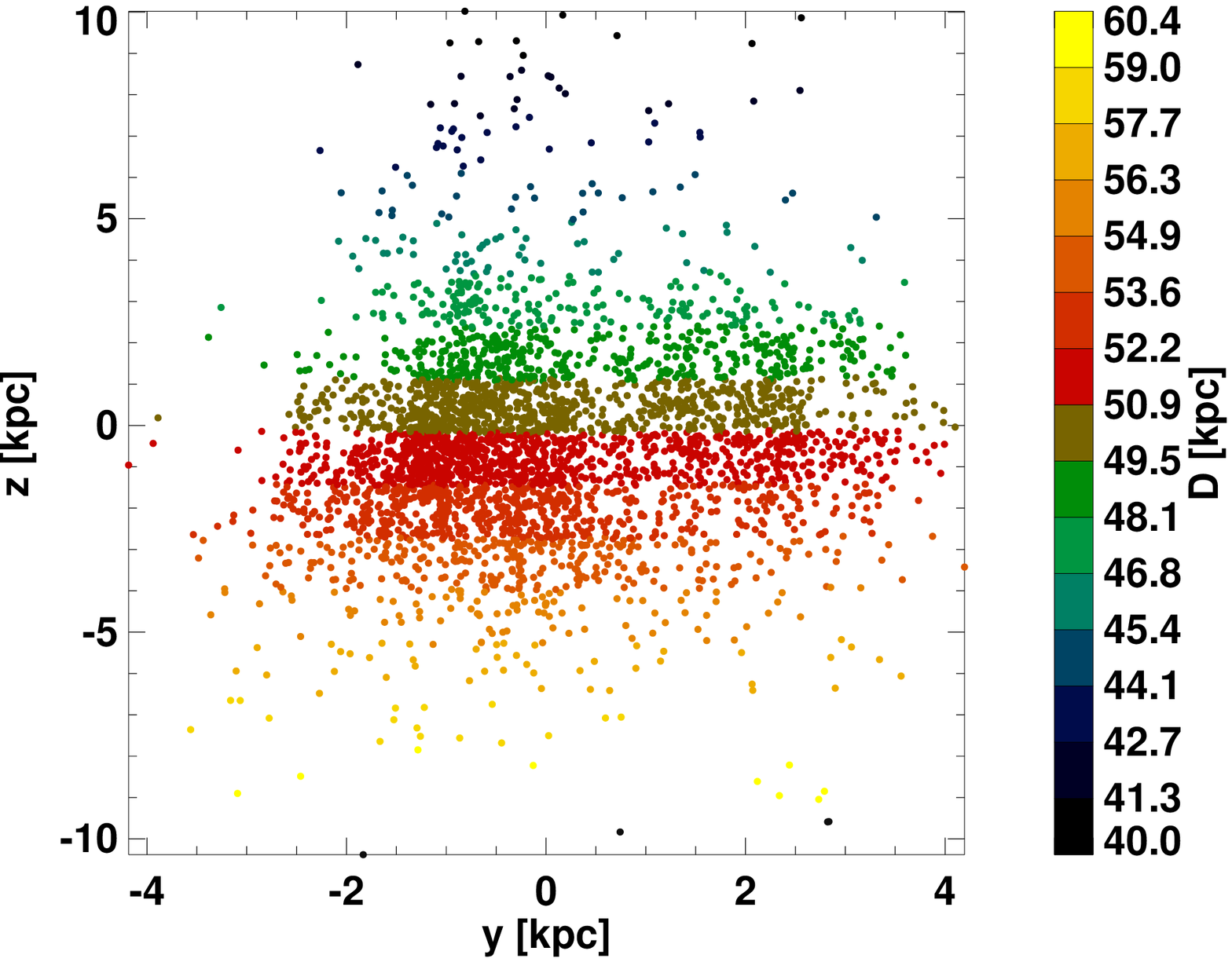}}&
  \resizebox{0.32\linewidth}{!}{\includegraphics*{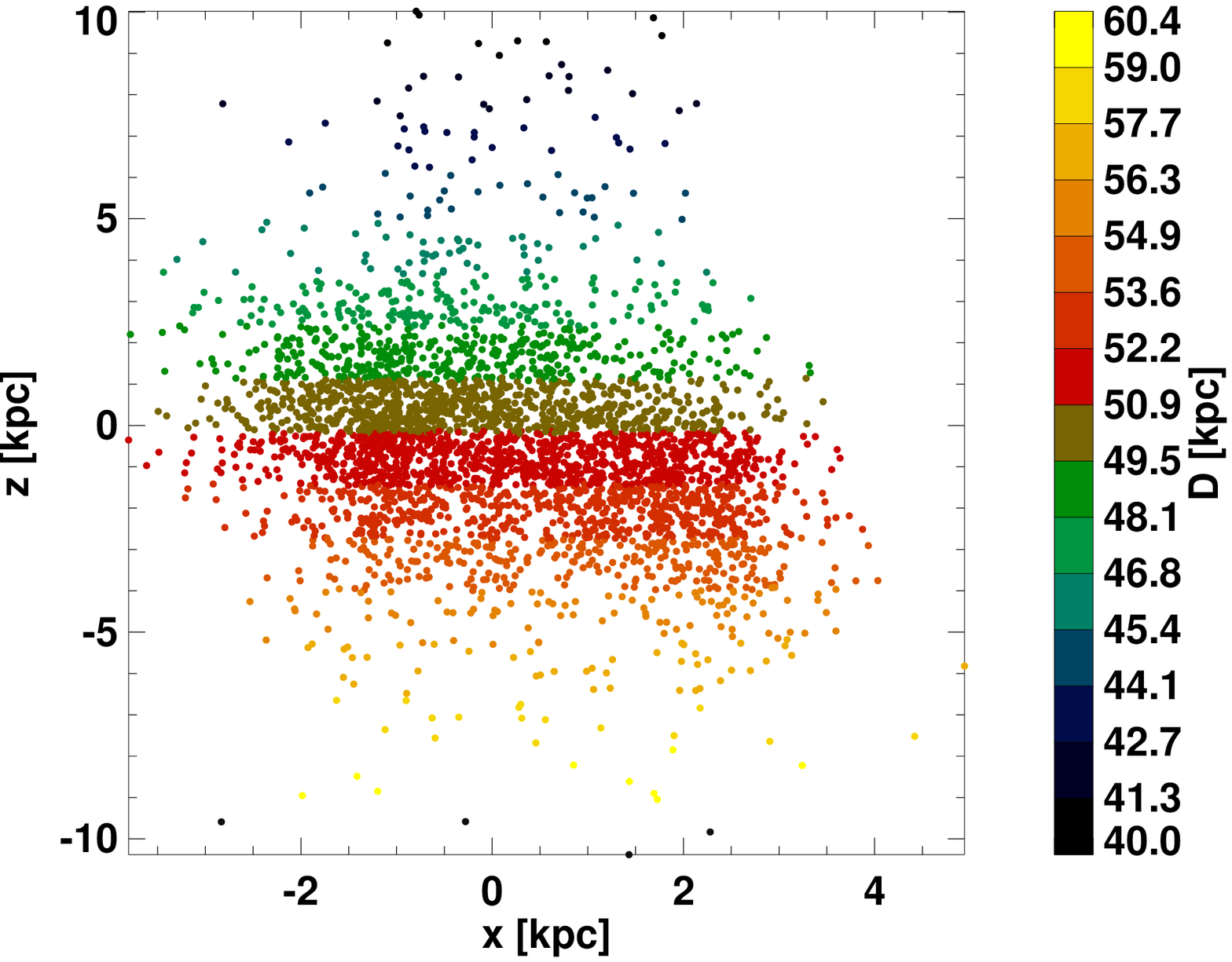}}\\
\vspace{-0.04\linewidth}
\end{tabular}
\caption{The projected distribution of LMC Cepheids on the  $XY$, $YZ$ and 
$XZ$ planes with the colour bar plots of their distances.}
\label{fig:xyz}
\end{figure*}

The \citet{inno16} catalog gives the ($\alpha,\delta$) values for individual 
stars, 
and the distances ($D$) are obtained through the period-luminosity relations, 
as described in Section \ref{sec:pl}. The values of these quantities 
$(\alpha,\delta,D)$ are transformed into the Cartesian coordinate system 
$(x, y, z)$, where the $z$-axis is pointed towards the observer, the $x$-axis 
is antiparallel to the $\alpha$-axis, and the $y$-axis is parallel to the 
$\delta$-axis. $D_0$ is the distance between the center of the target galaxy 
and the observer, while $D$ is the distance to a
particular star, and ($\alpha_0$, $\delta_0$) is the center of the target galaxy
in equatorial coordinates. The center of the LMC in the present study is taken
as $(\alpha_{0},\delta_{0})=(80^{\circ}.78,-69^{\circ}.03)$ \citep{niko04} and 
the distance to the center is taken as $D_{0}=49.973$ kpc \citep{piet13}. The 
transformation from $(\alpha, \delta, D)$ to $(x, y, z)$ can be performed by 
the transformation equations
\begin{align}
  \label{eq:radec2xyz}
  x
  &=
  -D \sin(\alpha - \alpha_0) \cos\delta,
  \notag \\
  y
  &=
  D \sin\delta \cos\delta_0 -
  D \sin\delta_0 \cos(\alpha - \alpha_0) \cos\delta,
  \notag \\
  z
  &=
  D_0 - D \sin\delta \sin\delta_0 - 
  D \cos\delta_0 \cos\alpha -
  \alpha_0 \cos\delta.
\end{align}
The coordinate system of the LMC $(x^{\prime}, y^{\prime}, z^{\prime})$ is the 
same as the coordinate system $(x, y, z)$, except that it has been rotated 
counterclockwise by an angle $\theta$ about the $z$-axis, and then clockwise 
by an angle $i$ about the new $x$-axis. These coordinate transformations can 
be written as
\begin{align}
\label{rmatrix}
\begin{bmatrix} x^{\prime} \\ y^{\prime} \\ z^{\prime} \end{bmatrix} 
=  \begin{bmatrix} 
\cos{\theta} & \sin{\theta} & 0  \\
-\sin{\theta}\cos{i} & \cos{\theta}\cos{i} & -\sin{i} \\
-\sin{\theta}\sin{i} & \cos{\theta}\sin{i} & \cos{i} 
\end{bmatrix}\begin{bmatrix} x\\ y\\ z \end{bmatrix}.
\end{align}
The Cartesian projected distributions of the LMC Cepheids are shown in 
Fig.~\ref{fig:xyz}. Let us denote the transformation matrix as 
\begin{align}
\label{rmatrix1}
T= & \begin{bmatrix} 
\cos{\theta} & \sin{\theta} & 0  \\
-\sin{\theta}\cos{i} & \cos{\theta}\cos{i} & -\sin{i} \\
-\sin{\theta}\sin{i} & \cos{\theta}\sin{i} & \cos{i} 
\end{bmatrix}.
\end{align}
 \subsection{Moment of Inertia Tensor Analysis}
\label{sub-sec:mita}
Distribution of Cepheids in the LMC  can be  modeled  as a rigid body whose  
inertial properties can be completely described by the inertia ellipsoid 
obtained analogously from the moment of tensor of rigid bodies. Let the 
reference coordinates of the sky plane be defined by Cartesian coordinates
$(x,y,z)$. The orientation of the galaxy with respect to the the plane of the 
sky can be described by an ellipsoid given by the rotated or transformed 
Cartesian coordinates $(x^{\prime},y^{\prime},z^{\prime})$ whose centroid 
coincides with the origin of the coordinate system ($x,y,z$). The transformed 
or rotated coordinate system $(x^{\prime},y^{\prime},z^{\prime})$ is obtained 
from the diagonalisation of the inertia tensor ${\mathbf I}$. The inertia 
tensor, $\mathbf{I}$, can be constructed  as follows \citep{deb14}:
\begin{equation}
  \label{eq:inertia-tensor}
  \mathbf{I} =
  \begin{bmatrix}
     I_{xx} & -I_{xy} & -I_{xz} \\
    -I_{yx} &  I_{yy} & -I_{yz} \\
    -I_{zx} & -I_{zy} &  I_{zz}
  \end{bmatrix}
\end{equation}
where the components are defined by
\begin{equation}
  \label{eq:inertia-components}
  I_{\alpha\beta} =
  \begin{cases}
    \sum_{\gamma \neq \alpha} Cov(\gamma, \gamma), & if \alpha = \beta
    \\
    Cov(\alpha, \beta), & if \alpha \neq \beta
  \end{cases}
\end{equation}
where covariance is defined by
\begin{equation}
  \label{eq:covariances}
  Cov(\alpha, \beta) =
  \frac{1}{N} \sum_{i=1}^N (\alpha(i) - \bar\alpha) (\beta(i) - \bar\beta),
\end{equation}
and the centroid is defined by
\begin{equation}
  \label{eq:mean}
  \bar\alpha = \sum_{i=1}^N \alpha(i) / N.
\end{equation}
The moment of inertia tensor ${\mathbf I}$ by construction itself is a 
symmetric matrix and hence can be easily diagonalised. The axes of the 
coordinate system $(x,y,z)$ are selected in such a way that 
$I_{xy}=I_{xz}=I_{yz}=0$ giving rise to the principal moments of inertia of the 
system given by $I_{xx},I_{yy}$ and $I_{zz}$, respectively. This can be 
obtained from the diagonalisation of the symmetric covariance matrix 
${\mathbf I}$ which is achieved through a  rotation or transformation matrix 
${\mathbf T}$. The matrix $\mathbf{T}$ is formed with the eigenvectors of 
$\mathbf{I}$ as its column vectors. The matrix $\mathbf{T}$ therefore serves 
as the rotation or  transformation matrix to the new coordinate system, 
$T: \mathbb{R}^3 \to \mathbb{R}'^3$. It can be seen that the matrix 
${\mathbf T}^{-1}\mathbf{I}\mathbf{T}$ is a diagonal matrix whose diagonal 
elements are the eigenvalues of the covariance matrix ${\mathbf{I}}$. The 
individual eigenvectors represent the three orthogonal axes in the new 
coordinate system $(x^{\prime},y^{\prime},z^{\prime})$, and are called the 
principal axes. Three eigenvalues ($\lambda_1 > \lambda_2 > \lambda_3$) 
correspond to these three eigenvectors, and are called the principal moments 
of inertia of the system. The principal axes with the smallest, intermediate 
and largest eigenvalues are often called the minor, intermediate and major 
axes, respectively of the inertia ellipsoid. The lengths of the semi-axes of 
the ellipsoid  are given by \citep{deb14}
\begin{align*}
S_{i}=& \sqrt{\frac{5}{2}\left(\lambda_{j}+\lambda_{k}-\lambda_{i}\right)},~~\text{for}~~i \ne j \ne k.
\end{align*}      
The transformation matrix or the rotation matrix ($\mathbf T$) of the 
transformation $T:R^{3} \rightarrow R^{\prime 3}$ obtained from the 
eigenvectors of the matrix ${\mathbf I}$ as column vectors is given by
\begin{align}
\label{tmatrix}
\begin{pmatrix} e_{1}^{\prime} \\ e_{2}^{\prime} \\ e_{3}^{\prime}\end{pmatrix}
=& \begin{pmatrix} 
T_{11} & T_{21} & T_{31} \\ 
T_{12} & T_{22} & T_{32} \\
T_{13} & T_{23} & T_{33} \\
\end{pmatrix} 
\begin{pmatrix}e_{1} \\ e_{2} \\ e_{3}\end{pmatrix},
\end{align}
where $T_{ij}=\cos{\left(e_{i}.e_{j}^{\prime}\right)}$ are the direction 
cosines. $\vec{e}=(e_{1},e_{2},e_{3})^{T}$ and  $\vec{e^{\prime}}=(e_{1}^{\prime},e_{2}^{\prime},e_{3}^{\prime})^{T}$ denote the basis vectors in the Cartesian 
coordinate system $(x,y,z)$ and the new rotated coordinate system 
$(x^{\prime},y^{\prime},z^{\prime})$, respectively. The transformation matrix
$\mathbf{T}$ contains the information about the orientation of the 
ellipsoid with respect to the unrotated coordinate system $(x,y,z)$ which 
can be used to find its position angle $(\theta)$ and inclination angle $(i)$
\citep{subr12,deb14}. It should be noted that the position angle $\theta$ as
defined in equation~(\ref{rmatrix}) is measured from the positive x-axis, 
i.e., from the west direction. However, in 
astronomical convention, position angles are 
always measured from the north ($0^{\circ}$) towards east ($90^{\circ}$) \citep{vand01}. Therefore, the value of $\theta$ as measured with the usual 
astronomical convention will be given by 
$\theta_{\text{lon}}=\theta+90^{\circ}$. Also, since the line of nodes is a 
line its value can be described by two different position angles that differ 
by $180^{\circ}$ \citep{vand01}.

Application of the moment of inertia tensor analysis to the $(x,y,z)$ 
distribution of the classical Cepheids without considering the errors $(\sigma_{x},\sigma_{y},\sigma_{z})$  yields the eigenvalues 
$\lambda_{1}=7.874,~\lambda_{2}
=7.224$ and $\lambda_{3}=4.182$. Using the transformation matrix constructed 
from the eigenvectors corresponding to these eigenvalues, the values of 
$\theta$ and $i$ are $15^{\circ}.858$ and $38^{\circ}.953$, respectively. 
However, the value of $\theta$ obtained here is measured 
with respect to the $x$-axis (west). If measured according to the astronomical 
convention from the positive $y$-direction (north), it comes out to be 
$\theta_{\text{lon}}= 128^{\circ}.953$. The final values of these parameters 
along with their errors are obtained using Monte Carlo simulations as 
described in \citet{deb15}. In order to carry out the Monte Carlo simulations, 
we have randomly generated Cartesian distributions $(x,y,z)$ using the errors
($\sigma_{x},\sigma_{y},\sigma_{z}$) from the observed distributions. The 
procedure of modeling the  randomly generated distribution of stars is  done
with the help of the  moment of tensor analysis for $10^{5}$ iterations. 
In each iteration of the Monte carlo simulations, the values of the parameters
of the ellipsoid such as $\{i,\theta_{{\rm lon}},S_{0},S_{1},S_{2}\}$ are 
noted down. Monte Carlo simulations for $10^{5}$ iterations helps to build up
the distribution of these five parameters. The histograms obtained from the 
individual distributions of each of the five parameters are fitted with a 
three-parameter Gaussian distribution which yield the respective values 
$\mu$, $\sigma$ and the peak of the histogram. From the Monte Carlo 
simulations, the following values of the parameters are obtained for the 
LMC with axes ratios $1.000\pm 0.003:1.151\pm0.003:1.890\pm 0.014$ and the 
viewing angle parameters:  inclination angle of $i=11^{\circ}.920\pm 0^{\circ}.315$ with respect to the longest axis from the line of sight and position angle 
of line of nodes $\theta_{\rm lon} = 128^{\circ}.871\pm 0^{\circ}.569$ as 
measured eastwards from north following astronomical convention. The lengths 
of the semi-major, intermediate and semi-minor axes for the galaxy are  
$S_{0}=5.712\pm 0.038$ kpc, $S_{1}=3.480\pm 0.003$ kpc, $S_{2}=3.021\pm0.008$ 
kpc, respectively, where $S_{0}>S_{1}>S_{2}$. 
\subsection{Plane fit solution to the $z$ distribution}
\begin{figure}
\begin{center}
\includegraphics[width=0.5\textwidth,keepaspectratio]{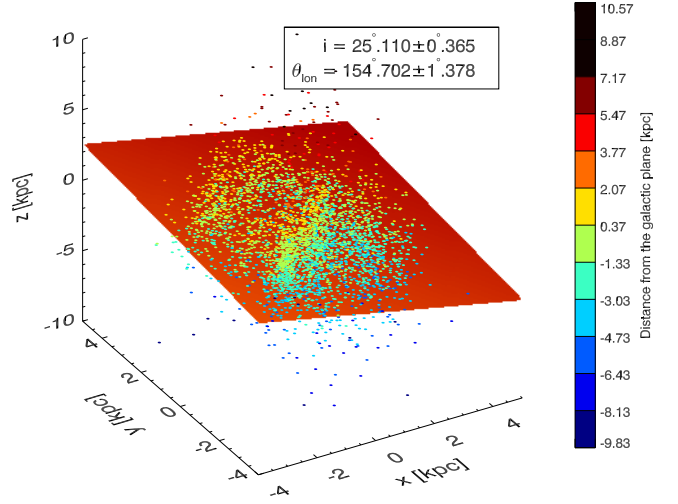}
\caption{Three dimensional ($x,y,z$) distribution of the combined (FO+FU)
sample of LMC classical Cepheids. The fitted plane with the parameters 
obtained from the plane fit is also overplotted.}
\label{fig:plane}
\end{center}
\end{figure}

Since we have the ($x,y,z$) distribution for each star in the Cartesian 
coordinate system, we now derive the inclination ($i$) and position angle 
($\theta_{\text{lon}}$) by fitting a plane solution of the following form 
\citep{niko04}:
\begin{align}
z=Ax+By+C.
\label{eq:plane}
\end{align}     
The values of $A$ and $B$ obtained from the plane fit solution given by 
equation~(\ref{eq:plane}) are then used to compute the position angle 
($\theta_{\text{lon}}$) and inclination ($i$) of the LMC disk:
\begin{align*}
\theta_{\text{lon}}=& {\rm arctan}\left(-\frac{A}{B}\right)+{\rm sign}(B)\frac{\pi}{2}, \\
i=& {\rm arccos}\left(\frac{1}{\sqrt{1+A^{2}+B^{2}}}\right).
\end{align*}
The standard errors in $\theta$ and $i$ are obtained from the propagation of 
error formula as follows:
\begin{align*}
\sigma_{\theta}= & \frac{1}{\sqrt{A^2+B^2}}\sqrt{A^{2}\sigma_{B}^2+B^{2}\sigma_{A}^2}, \\
\sigma_{i}= & \frac{1}{A^2+B^2+1}\frac{1}{\sqrt{A^2+B^2}}\sqrt{A^{2}\sigma_{A}^2+B^{2}\sigma_{B}^2}. 
\end{align*}
The results obtained from the plane fit solution are listed in 
Table~\ref{table:angle} using FO, FU and (FO+FU) Cepheids, respectively. The 
fitted plane using the parameters obtained from the plane fit  using (FO+FU) 
Cepheids is shown in Fig.~\ref{fig:plane}. We find that the values of the 
viewing angle parameters obtained by \citet{inno16} are in good agreement with
those obtained in the present study although the methodology of obtaining 
their values in these two studies are completely different. However, we note 
that the statistical errors of the derived viewing angle parameters as given 
in  \citet{inno16} are unrealistically small and  seem to be wrongly quoted. 
The errors quoted should actually be in radians if they are correctly 
calculated in \citet{inno16} and need to be multiplied by $\frac{180}{\pi}$ to 
convert them into degrees while writing down the results of the measured 
values. Making this correction, we write down the values obtained by 
\citet{inno16} as $i=25^{\circ}.05\pm 1^{\circ}.146$, $\theta_{\text{lon}}=150^{\circ}.76\pm 1^{\circ}.146$ . Comparing these values with those obtained in the
present study (third row of Table~\ref{table:angle}) show that the 
$\theta_{\text{lon}}$ value differs at the $2.2\sigma$ level and the $i$ values
are consistent within $0.2\sigma$. On the other hand, the values of 
$i=26^{\circ}.2\pm 5^{\circ}.9$ and $\theta_{\text{lon}}=154^{\circ}.5\pm2^{\circ}.1$ found by \citet{vand14} combining the proper motion measurements and 
line-of-sight velocities for young stars in the LMC are consistent with their 
values obtained in the present study within $0.2\sigma$ and 
$0.1\sigma$ levels, respectively. The viewing angle parameters obtained in 
the present study are also comparable to the values of 
$i=23^{\circ}.5\pm 0^{\circ}.4$ and 
$\theta_{\text{lon}}=154^{\circ}.6\pm 1^{\circ}.2$ obtained by \citet{koer09} 
using the Red Clump stars. A comprehensive list of the viewing angle 
parameters for the LMC available in the literature obtained using various 
tracers can be found in \citet{inno16}.                  

One can easily see that the values of $(i,\theta_{\text{lon}})$ obtained from
the moment of tensor analysis and the plane fitting procedure differ 
significantly because these two methods are based on completely different 
principles. Most of the values of the viewing angle parameters of the LMC as
quoted in the literature are based on the plane fitting procedure and hence 
the comparisons of these values have been done obtained using this method 
in the present study. However the moment of tensor analysis yields the values 
of the axes ratios of the LMC apart from finding the viewing angle parameters. 
This is particularly useful to know the extent of the LMC in three 
perpendicular directions.         
\begin{table*}
\begin{center}
\caption{Plane fit solutions to the $z$-distribution of the LMC Cepheids as a 
function of ($x,y$) coordinates using FO, FU and (FO+FU) Cepheids. The 
resulting viewing angle parameters, $i$ and $\theta$ calculated from solutions
of the plane fit.}
\label{table:angle}
\begin{tabular}{lcccc} \\ \hline
Solution & $i$ & $\theta_{\text{lon}}$& $\chi^{2}$ & Remarks \\ \hline \hline 
$z_{\rm fit}=(-0.432\pm0.016)x+(0.233\pm 0.015)y+(-0.193\pm 0.023)$ & $26^{\circ}.151\pm 0^{\circ}.499$ & $151^{\circ}.614\pm1.764$ &$3.884$ & FO only \\
$z_{\rm fit}=(-0.410\pm0.017)x+(0.156\pm 0.017)y+(-0.205\pm 0.025)$ & $23^{\circ}.691\pm 0^{\circ}.534$ & $159^{\circ}.206\pm2.229$ &$1.662$ & FU only \\
$z_{\rm fit}=(-0.424\pm 0.011)x+(0.200\pm 0.011)y+(-0.195\pm 0.017)$ & $25^{\circ}.110\pm 0^{\circ}.365$ & $154^{\circ}.702\pm1.378$ &$2.586$ & (FO+FU) \\ \hline
\end{tabular}
\end{center}
\end{table*}
\section{Separation of Bar and disk of the LMC}
\label{sec:bar_disk}
The LMC has a prominent luminous central optical bar which is slightly 
off-centered and misaligned with the disk \citep{zhao00}. The bar is 
morphologically peculiar in the sense that the eastern tip of the bar shows
an overdensity and is dominated by the patchy distribution of young stars 
and higher values of dust extinction as seen in optical images \citep{vand14}. 
The bar of the LMC has been observed in the distributions of young, 
intermediate-age and and old stellar populations. The bar is roughly of the 
$(3^{\circ}\times 1^{\circ})$ angular size \citep{west97, manc04}. A number of 
numerical simulations have been carried out in the 
recent past to explain the formation of the off-center bar of the LMC. It was
suggested by \citet{zhao00} that the off-center LMC bar is the result of a 
recent tidal interaction of the LMC with the SMC and the Milky Way.          
In one of the studies,  the off-center bar of the LMC is shown to be the 
result of collision of the LMC with a low-mass sub-halo (dark satellites) of
mass $10^{8}-10^{9}~M_{\odot}$ belonging to the Galaxy from numerical 
simulations \citep{bekk09}. 
The most recent studies attempt to explain the formation of the bar due to the 
close encounter of the LMC and SMC (dwarf-dwarf galaxy interaction) a few Myr 
ago \citep{besl12,yozi14,pard16,kruk17}. Using a sample of $270$ late type 
galaxies from the multi-wavelength Sloan Digital Sky Survey (SDSS) images and 
Galaxy Zoo morphologies, \citet{kruk17} measured offsets in the range $0.2-2.5$
kpc between the photometric centres of the stellar disk and stellar bar. 
Majority of the galaxies with off-centre bars in that study are of Magellanic 
type and this provides evidences in the support of predictions from the
simulations of dwarf-dwarf galaxy interactions. Reconstructing the observed
structures of a galaxy from the simulations of galaxy interactions is often
quite challenging and provides an insight into the understanding of the 
evolution of the Universe.

\begin{figure}
\begin{center}
\includegraphics[width=0.5\textwidth,keepaspectratio]{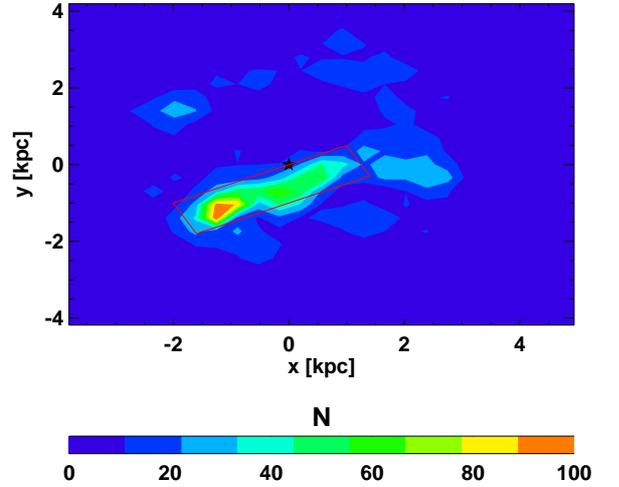}
\caption{Two dimensional density contours of the LMC. The bar is shown as the 
rectangular region in the $XY$-plane satisfying the inequality given by 
equation~(\ref{eq:bar}).}
\label{fig:contour}
\end{center}
\end{figure}

\begin{figure}
\begin{center}
\includegraphics[width=0.5\textwidth,keepaspectratio]{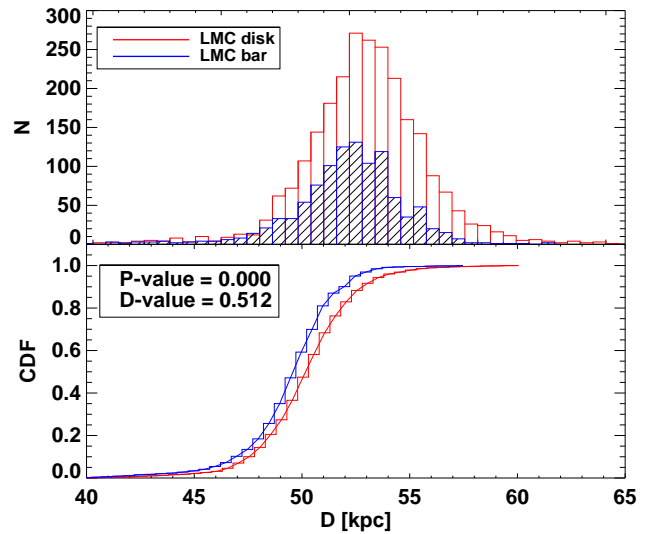}
\caption{Cumulative distributions of line of sight distances of disk and 
bar Cepheids are shown as red and blue solid lines, respectively. The bar 
and disk Cepheids are selected whether the inequalities defined in 
equation~(\ref{eq:bar}) are satisfied or not.}
\label{fig:cumulate_bar}
\end{center}
\end{figure}

From the study of three dimensional structure of the LMC using different 
tracers, a number of studies were devoted to finding out whether the bar of the 
LMC is off-centred and misaligned with respect to its disk 
\citep{niko04,koer09,subr13,vand14,jacy16}. All these studies except the one
by \citet{jacy16} found the bar of the LMC to be offset from its disk with the 
former being closer to the observer than the latter. In the the study of three
dimensional structure of the LMC using OGLE classical Cepheids, \citet{jacy16}
redefined the definition of the bar stating that the entire bar should consist 
of the eastern part as well as extend towards the western end of the LMC 
contrary to the existing conventional definition of the bar in the literature 
which is believed to be consisting of only the eastern part of the LMC. 
Nonetheless, we 
select the bar region following its conventional definition as in the 
literature. In order to get an idea about the LMC bar and disk from the 
present analysis carried out in seven bands using a statistically larger 
sample of classical Cephieds, we make a slight 
modification of the definition of the bar following \citet{niko04}. This is 
done just to ensure that the dimension of the bar is roughly of the order of 
$\approx (3~\text{kpc}\times 1~\text{kpc})$ in the $XY$-plane as viewed along 
the line of nodes noting that $1^{\circ} \sim 1~\text{kpc}$ at the distance of 
the LMC. The bar is defined as the rectangular region in the $XY$- plane using 
the inequalities
\begin{align}
~~~~~~~~~~~~~~~~~~~~~~~&y  > 0.5x-1.0 \nonumber \\
~~~~~~~~~~~~~~~~~~~~~~~&y  < 0.5x \nonumber \\
~~~~~~~~~~~~~~~~~~~~~~~&y  > -2x-5.0 \nonumber \\
~~~~~~~~~~~~~~~~~~~~~~~&y  < -2x+2.5
\label{eq:bar}
\end{align}  
Stars satisfying the above inequalities are designated as ``bar" Cepheids 
else as ``disk" Cepheids. The location of the bar is overplotted on the
two dimensional density contour map of the LMC in Fig.~(\ref{fig:contour}).     
Once the separation between the disk and bar 
Cepheids is made, we shift the bar and the disk towards the observer along 
the line of nodes through a clockwise rotation about the $z$ axis according 
to the following transformation 
\begin{align}
x^{\prime}= & x\cos{\theta^{\prime}}+y\sin{\theta^{\prime}} \nonumber \\
y^{\prime}=& -x\sin{\theta^{\prime}}+y\cos{\theta^{\prime}} \nonumber  \\
z^{\prime}=& z,
\end{align}
where $\theta^{\prime}=\pi-\theta_{\text{lon}}$. A plane fit solution of the 
form  same as equation~(\ref{eq:plane}) using the deprojected Cartesian 
coordinates $(x^{\prime},y^{\prime},z^{\prime})$ obtained above is applied 
separately for
the disk and bar Cepheids. From the coefficients of the plane fits, we find 
that the inclinations and position angle of lines of nodes of the disk and bar 
are $(25^{\circ}.724\pm0^{\circ}.392,148^{\circ}.060\pm 1^{\circ}.444)$ and 
$(46^{\circ}.169\pm3^{\circ}.652,113^{\circ}.303\pm 4^{\circ}.882)$, 
respectively, with an offset of $1.182\pm 0.082$ kpc between the two, the bar 
being closer to us by $\sim 1$ kpc from the disk of the LMC. These results are 
in good agreement with the postulates put forwarded by \citet{zhao00} in order 
to explain the microlensing optical depth observed towards the LMC. In the 
model proposed by 
\citet{zhao00}, it was postulated that the disk and bar stars of the LMC are 
slightly misaligned with different inclinations having an offset of $\sim 1$ kpc
along the line of sight. Based on relative distance measurements of more than 
$2000$ Cepheids, \citet{niko04} found that the bar of the LMC is located
$\sim 0.5$ kpc in front of the main LMC disk, where it was suggested as a lower 
limit on the actual offset. Using the magnitudes of Red Clump
stars, \citet{koer09} also found the bar to be floating over the disk by 
$\sim1$ kpc and is nearer to us. \citet{vand14} also obtained the dynamical 
center of the rotating HI disk of the LMC to be $\ge 1$ kpc away from the 
densest point in the bar. These findings are in excellent agreement with the 
results obtained in the present study. The existence of an offset between the 
disk and bar of the LMC  is further substantiated using the two sample 
Kolmogorov-Smirnov (KS) statistics. The KS test allows to reject or confirm 
the hypothesis whether the two samples (data sets) have a common parent 
distribution. The two sample KS test returns a distance statistics $D$ defined
as the maximum distance between the cumulative distribution functions (CDFs) 
of the two samples along with the corresponding probability $P(D)$ if the two 
samples are drawn from the same parent distribution or not. A probability 
with  $P(D) < 10\%$ indicates that the two distributions are significantly 
different \citep{pres02}. CDFs of line of sight distances of disk (sample 1) 
and bar (sample 2) Cepheids are shown in Fig.~\ref{fig:cumulate_bar}. In order 
to check whether the two samples came from the same distribution at a 
significance level of $0.001$ we use the two sample KS statistics. The KS test 
applied on the distance CDFs yields $D$-value=0.512 with $P$-value=$0.000$. 
These values indicate the CDFs of the two samples are different at a 
significance level of $0.001$ which implies that the LMC bar region is closer 
to us than its disk. Not only there exists differences in the structural 
parameters between the LMC bar and disk as found in the present study, but 
also subtle differences in the $[\alpha/Fe]$ values between these two 
components of the LMC were confirmed by \citet{vand13}. These findings 
reflect the fact that bar is not merely an overdensity but related to a 
fresh episode of star formation in the central parts of the LMC around 
$\sim 5$ Gyr ago while the star formation history of the bar and inner disk 
were similar at earlier epochs \citep{vand13}.

\begin{figure*}
\vspace{0.02\linewidth}
\begin{tabular}{cc}
\vspace{+0.01\linewidth}
  \resizebox{0.45\linewidth}{!}{\includegraphics*{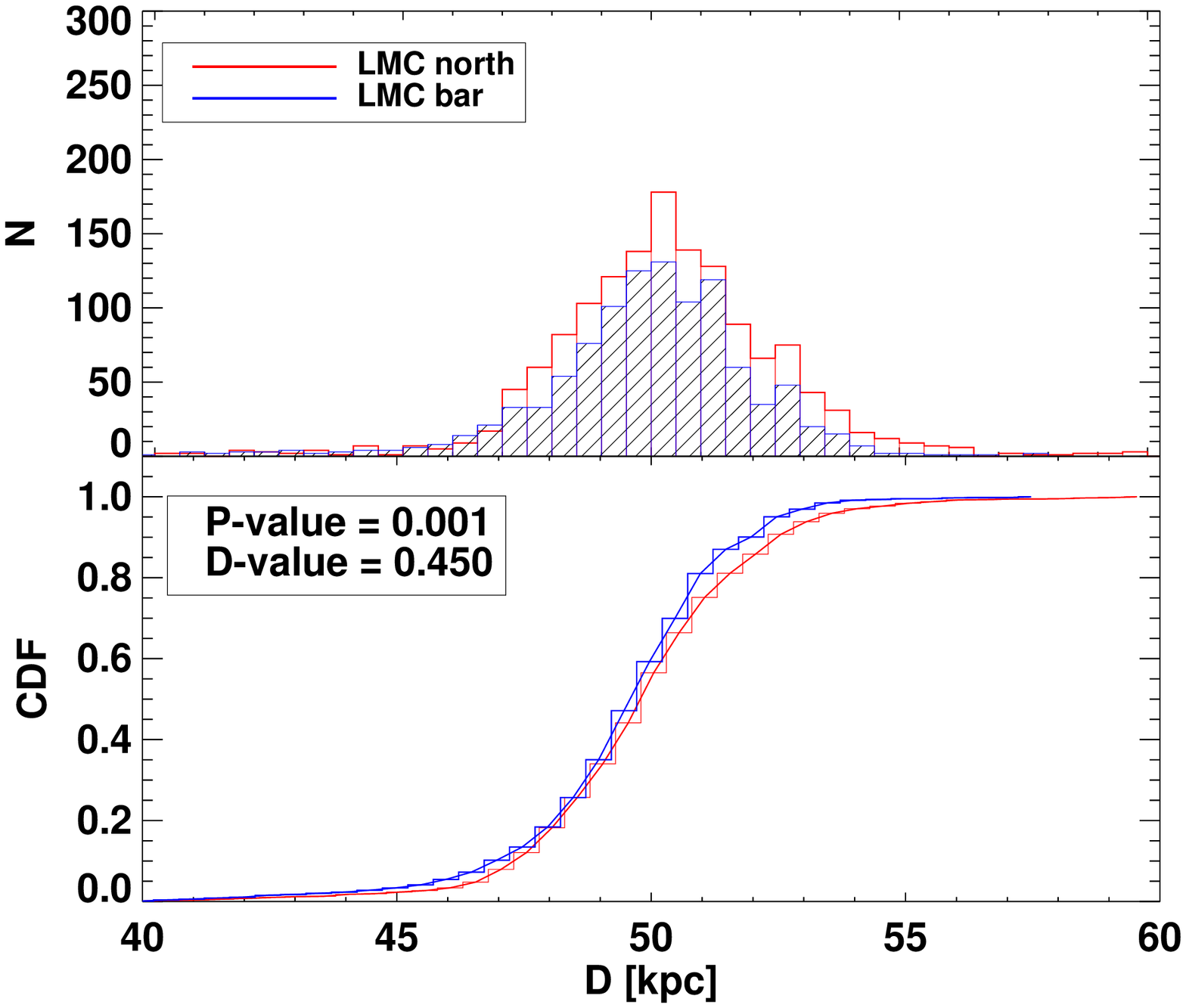}}&
\resizebox{0.45\linewidth}{!}{\includegraphics*{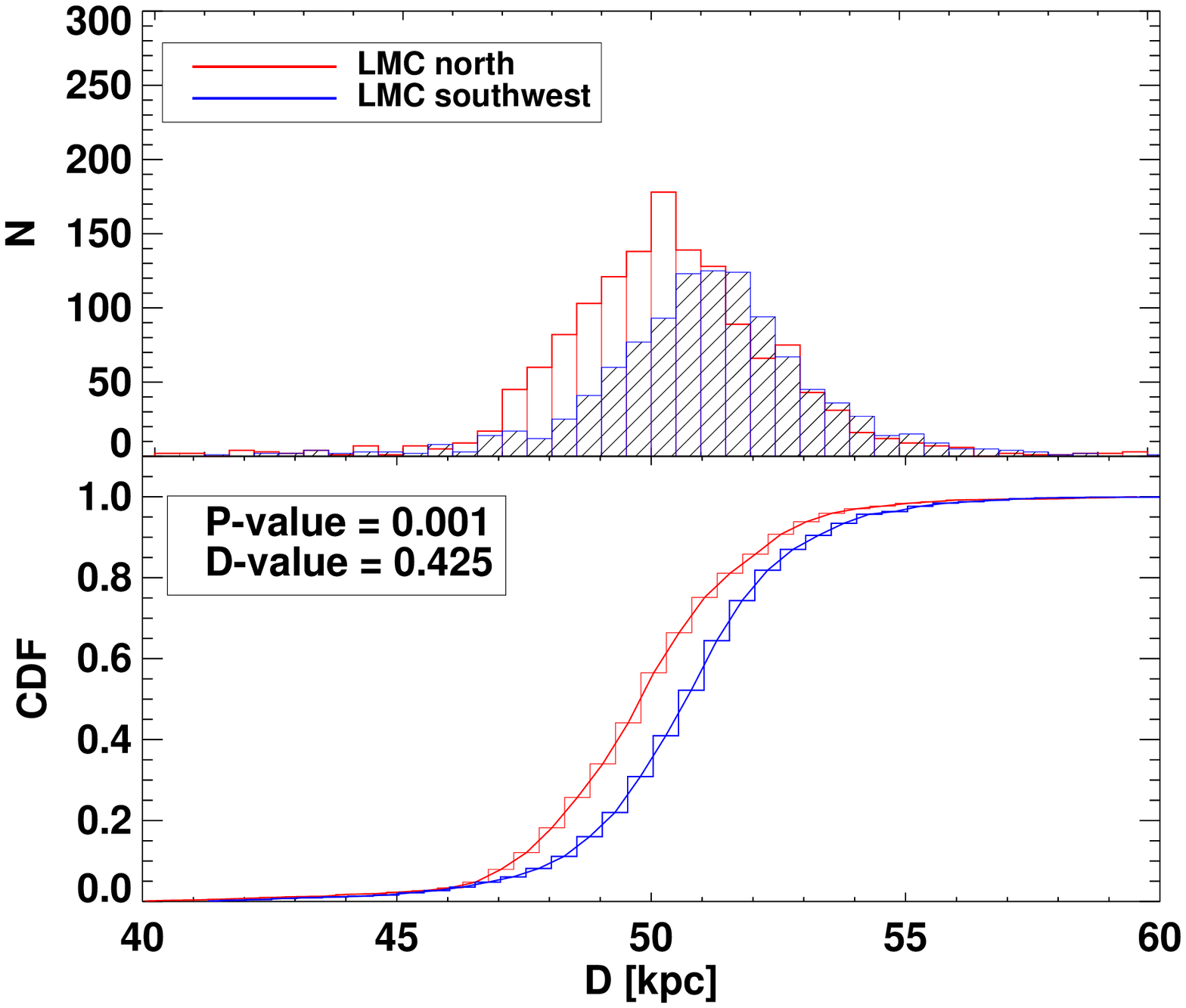}} \\ 
\vspace{-0.04\linewidth}
\end{tabular}
\caption{CDFs of line of sight distances between the bar and the northern part
of the LMC disk as well as between the northern and the southwestern parts
of the LMC disk are shown in the left and right panels, respectively.}
\label{fig:cumulate_others}
\end{figure*}

After finding that the bar of the LMC is nearer to us than the entire disk 
of the LMC, we now compare the CDFs between (i) the bar and the northern part 
of the LMC disk ($y>0$ and excluding the bar region) (ii) northern part and 
southwestern part of the LMC disk ($y<0$ and excluding the bar region). The 
CDFs of the comparison between these two different regions are shown in 
Fig.~\ref{fig:cumulate_others}. The KS test performed on these CDFs show that
the bar of the LMC lies at a closer distance than the northern part of the 
disk whereas, the southwestern parts of the LMC disk are located at the 
farthest distance of the LMC from the line of sight of a terrestrial observer. 
Similar test performed on the western and southern parts of the LMC disk 
indicates that southern part is situated more nearer to us than its 
western part.  
\section{Summary and Conclusions}
\label{sec:summary}
In this paper, we have utilised the mutli-wavelength  archival data of more 
than $3500$ common classical Cepheids available in seven photometric bands, 
viz., $V,I,J,H,Ks,[3.6]~\mu m$ and $[4.5]~\mu m$ to determine the geometrical 
and viewing angle parameters of the LMC. Solving the PL relations 
simultaneously in these bands, we obtain the following results:
\begin{enumerate}
\item Relative values of reddening and true distance moduli for individual
classical Cepheids with respect to the mean values of reddening and distance 
modulus of the LMC. In order to determine these values, apparent distance 
moduli calculated using the mean magnitudes and the PL relations obtained in 
each of the seven photometric bands are fitted simultaneously as a 
function of inverse wavelength $(1/\lambda)$ using the \citet{card89} 
reddening law.
\item The PL relations are obtained in two iterations. In iteration 1, these
relations are obtained without any reddening and distance corrections.       
In iteration 2, the PL relations are obtained by correcting the relative 
distance and reddening values obtained using the relations found in 
iteration 1. The dispersions in the PL relations obtained from iteration 2 
decrease substantially due to the distance and reddening corrections leading 
to the much more improved relations. The small residual dispersions in the 
improved PL relations exist due to the intrinsic error accounting for the 
finite width of the instability strip  as well as photometric errors 
corresponding to different photometric bands. Since the amplitudes of the 
light curves of Cepheids are a function of wavelengths and decrease while 
going from optical to mid-infrared photometric bands, the photometric errors increase 
accordingly. This is reflected in the resulting PL relations as we go from 
optical to mid-infrared photometric bands.            
\item The reddening and distance modulus offsets obtained from the simultaneous
fitting of apparent distance moduli in seven photometric bands are converted 
into their absolute values of reddening and true distance modulus using the 
mean values of reddening and distance modulus of the LMC taken from the 
literature.  
\item The reddening map of the LMC is constructed from the absolute reddening 
values distributed in the $(\alpha,\delta)$ coordinates. Two prominent regions 
having higher values of reddening can be identified from the map. One of them 
is the $30$ Dor region which has the highest reddening values. The other one 
with higher values of reddening can be associated with the LMC HI supergiant 
shells SGS (LMC). The reddening map constructed in this paper is found to be in 
excellent agreement with those found by \citet{niko04} and \citet{inno16}, 
respectively.     
\item The values of distance along with the information of equatorial 
coordinates ($\alpha,\delta$) are used to convert them into the corresponding
Cartesian coordinates with respect to the plane of the sky. 
By fitting a plane solution of the form $z=f(x,y)$ to the observed three 
dimensional distribution of the LMC Cepheids, we find the following viewing 
angle parameters: inclination angle $i=25^{\circ}.110\pm 0^{\circ}.365$ and 
position angle of line of nodes $\theta_{\text{lon}}=154^{\circ}.702\pm 1^{\circ}.378$.  On the other hand, modelling the observed three dimensional 
distribution of the Cepheids as a triaxial ellipsoid, the following values of 
the geometrical axes ratios of the LMC are obtained: $1.000\pm 0.003:1.151\pm0.003:1.890\pm 0.014$ with the viewing angle parameters:  inclination angle of 
$i=11^{\circ}.920\pm 0^{\circ}.315$ with respect to the longest axis from the 
line of sight and position angle of line of nodes $\theta_{\rm lon} = 128^{\circ}.871\pm 0^{\circ}.569$ as measured eastwards from north following 
astronomical convention.           
\item Separating the bar and disk of the LMC, we find that the bar and disk 
have significantly different values of position angles of line of nodes as well 
as inclinations, i.e. they are misaligned with respect to each other. An 
offset of $\sim 1$ kpc has also been obtained between the two with the bar 
being closer to us than the disk. Higher value of inclination angle found for 
the bar also corroborates the fact that it is closer to us than the disk. 
These values obtained from the precise multi-wavelength observational data of 
classical Cepheids provide evidences in support of the postulates of the model 
proposed by  \citet{zhao00} as an explanation of the microlensing optical 
depth observed towards the LMC.      
\end{enumerate}               
\section*{Acknowledgments}     
The authors acknowledge the use of highly valuable publicly available 
OGLE-IV data for this study. SD thanks Science and Engineering 
Research Board (SERB), Department of Science \& Technology (DST), Govt. of 
India for financial support of this study through a research grant D.O No. 
$\text{SB/FTP/PS-029/2013}$ under the  Fast Track Scheme for Young 
Scientists in Physical Sciences. CCN thanks the funding from Ministry of Science and Technology (Taiwan) under the contract 104-2112-M-008-012-MY3. HPS and 
SMK acknowledge support from Indo-US Science \& Technology Forum (IUSSTF) 
through the Indo-US virtual networked center on ``Theoretical Analyses of 
Variable Star Data". Lastly, the authors thank the anonymous referee for 
making various helpful comments and useful suggestions which made the paper 
significantly relevant. The paper makes use of the facility from 
arxiv.org/archive/astro-ph, NASA ADS and SIMBAD databases. 
\bibliographystyle{mnras}
\bibliography{deb}
\bsp
\label{lastpage}
\end{document}